\documentclass[sigconf]{acmart}
\usepackage{xspace}
\usepackage{xcolor}
\usepackage{paralist}
\usepackage{wrapfig}
\usepackage{multirow}
\usepackage{listings}
\usepackage{makecell}
\usepackage{boxedminipage}
\usepackage{booktabs}
\usepackage{subcaption}
\usepackage{tabularx}
\usepackage{url}
\usepackage{amsmath}

\usepackage{breakurl} 
\usepackage{boldline}
\usepackage[ruled,vlined]{algorithm2e}
\usepackage[titletoc]{appendix}

\usepackage{amssymb}%
\usepackage{pifont}%
\usepackage{balance}
\usepackage[htt]{hyphenat}

\AtBeginDocument{%
  }

\setcopyright{acmlicensed} %
\copyrightyear{2018} %
\acmYear{2018} %
\acmDOI{XXXXXXX.XXXXXXX} %
\acmConference[Conference acronym 'XX]{Make sure to enter the correct
  conference title from your rights confirmation email}{June 03--05,
  2018}{Woodstock, NY}  %
\acmISBN{978-1-4503-XXXX-X/2018/06}  %


\newcommand{\bheading}[1]{{\vspace{1pt}\noindent{\textbf{#1}}}}
\newcommand{\iheading}[1]{{\vspace{1pt}\noindent{\textit{#1}}}}

\newcommand{\cc}[1]{\mbox{\smaller[0.5]\texttt{#1}}}

\def\Snospace~{\S{}}

\newcounter{note}[section]

\newcommand{\ycrv}[1]{\textcolor{black}{{#1}}}

\newcommand{\etal}{\emph{et al.}\xspace}
\newcommand{\etc}{\emph{etc}\xspace}
\newcommand{\ie}{\emph{i.e.}\xspace}
\newcommand{\eg}{\emph{e.g.}\xspace}

\newcommand{\ignore}[1]{}

\newcounter{packednmbr}

\newenvironment{packeditemize}{
\begin{list}{$\bullet$}{
\setlength{\labelwidth}{8pt}
\setlength{\itemsep}{0pt}
\setlength{\leftmargin}{\labelwidth}
\addtolength{\leftmargin}{\labelsep}
\setlength{\parindent}{0pt}
\setlength{\listparindent}{\parindent}
\setlength{\parsep}{0pt}
\setlength{\topsep}{3pt}}}{\end{list}}

\newcommand{\sysname}{\textsc{AVP-Inspect}\xspace}
\newcommand{\compone}{\texttt{Controller}\xspace}
\newcommand{\comptwo}{\texttt{Explorer}\xspace}
\newcommand{\compthree}{\texttt{Detector}\xspace}

\copyrightyear{2026}
\acmYear{2026}
\setcopyright{cc}
\setcctype{by}
\acmConference[CCS '26]{Proceedings of the 2026 ACM SIGSAC Conference on Computer and Communications Security}{November 15--19, 2026}{The Hague, Netherlands}
\acmBooktitle{Proceedings of the 2026 ACM SIGSAC Conference on Computer and Communications Security (CCS '26), November 15--19, 2026, The Hague, Netherlands}
\acmDOI{10.1145/3830454.3846586}
\acmISBN{979-8-4007-2871-6/2026/11}

\begin{document}

\title{\sysname: Coordinated Cyber-Physical Testing for Privacy Analysis of COTS Apple Vision Pro Applications}

\author{Yichang Xiong}
\affiliation{%
  \institution{George Mason University}
  \city{Fairfax}
  \state{Virginia}
  \country{USA}
}
\email{yxiong2@gmu.edu}
 
\author{Vamsi Shankar Simhadri}
\affiliation{%
  \institution{George Mason University}
  \city{Fairfax}
  \state{Virginia}
  \country{USA}
}
\email{vsimhadr@gmu.edu}
 
\author{Yue Xiao}
\affiliation{%
  \institution{William \& Mary}
  \city{Williamsburg}
  \state{Virginia}
  \country{USA}
}
\email{yxiao05@wm.edu}
 
\author{Xiaokuan Zhang}
\affiliation{%
  \institution{George Mason University}
  \city{Fairfax}
  \state{Virginia}
  \country{USA}
}
\email{xiaokuan@gmu.edu}

\begin{abstract}
XR devices introduce substantial privacy concerns due to their comprehensive data collection capabilities that surpass traditional computing platforms.
While existing works have demonstrated privacy concerns on Android-based XR devices such as Meta Quest series by performing network traffic analysis, little attention has been paid to the Apple Vision Pro (AVP) devices,
mainly due to the closed nature and the technical challenges associated with AVP devices.
In this work, we make a bold attempt to detect privacy violations of AVP applications from network traffic through automatic testing on AVP devices.
Our key insight is that effective AVP application testing requires coordinated control of both cyber (software) and physical (hardware) components,
which we term \textit{Coordinated Cyber-Physical Testing}.

Building on this insight,
we design and implement \sysname, an automatic dynamic analysis framework for AVP applications,
overcoming significant challenges enforced by the closed-source nature of AVP ecosystem.
\sysname consists of three components: 
an automatic device controller by building customized hardware devices,  a 3D UI explorer by designing a new exploration engine, and a privacy violation detector by constructing a unified privacy taxonomy for AVP.
We first evaluated \sysname on a manually constructed ground truth dataset, 
then performed a large-scale analysis on 324 AVP applications downloaded from the App Store,
with each app tested for 20 minutes.
{We found that 188 (58.0\%) of apps exhibit at least one violation, and more than 60\% of the network traffic flows are not properly disclosed.}

\end{abstract}

\maketitle

\section{Introduction}
\label{sec:intro}

Extended Reality (XR) technology is emerging as a fundamental computing paradigm that merges physical and digital worlds, with market projections reaching \$472.39 billion by 2029~\cite{mr-market}. XR encompasses mixed reality (MR), which combines augmented reality (AR) and virtual reality (VR) to enable real-time interactions between physical and virtual objects. This integration creates immersive environments that enhance user interaction and experience.
XR devices have demonstrated practical value across multiple domains, including healthcare, education, entertainment, and professional training~\cite{vr-domain-edu-1,vr-domain-health-2,vr-domain-military-1,fire3}.
The recent introduction of Apple Vision Pro (AVP)~\cite{avp-web} represents a significant development in the XR landscape, potentially accelerating mainstream adoption in both consumer and enterprise markets.
Recent deployments of XR devices for surgical assistance~\cite{avp-surgery2}
further demonstrate XR's emergence as a revolutionary computing platform that fundamentally transforms human-technology interaction.

Although XR devices bring a lot of benefits,
they introduce substantial privacy concerns due to their comprehensive data collection capabilities that surpass traditional computing platforms. These devices continuously gather extensive biometric and behavioral data, including eye movements, facial expressions, hand gestures, and full body movements. Recent research has demonstrated how this data can be exploited to extract sensitive information such as passwords, personal characteristics, and health conditions~\cite{zhang_its_2023,wu_privacy_2023,luo_holologger_2022,slocum_going_2023,meteriz-yildiran_keylogging_2022,cayir2025speak,ye2024bpsniff}.
Moreover, XR applications must comply with privacy regulations such as the General Data Protection Regulation (GDPR)~\cite{GDPR} and California Consumer Privacy Act (CCPA)~\cite{CCPA}, by transparently and correctly documenting their data collection, processing, and sharing procedures through comprehensive privacy policies.

While existing research has made important progress in understanding privacy risks in popular XR platforms such as Meta's Quest series~\cite{trimananda2022ovrseen,zhan2024vpvet}, their methods cannot be adapted to AVP due to significant differences in both the hardware and software stacks.
The closed-source nature of commercial XR platforms presents significant technical barriers to privacy analysis. 
This technical obstacle is especially pronounced with AVP due to Apple's restrictive policies regarding third-party system analysis. Unlike iOS, which has benefited from two decades of community-driven security research and analysis tools (\eg, jailbreaking techniques for root access~\cite{ios-jailbreak1} and dynamic analysis tools~\cite{ios-frida,ios-dynamic-1}), AVP was released in early 2024 and lacks established security testing frameworks. 
Consequently, there exists no comprehensive privacy analysis of the AVP platform, leaving critical questions about its privacy protections unanswered.

\bheading{Goals and challenges.}
This research aims to make a bold attempt to bridge the critical research gap: \textit{detecting privacy  non-compliance in commercial-off-the-shelf (COTS) AVP applications through network traffic analysis}.
However, constructing an effective dynamic analysis framework faces three key technical challenges.
\begin{packeditemize}
\item \textbf{C1:} AVP's closed system architecture prevents automated testing. Unlike established platforms such as Android that provide debugging interfaces and root access capabilities, AVP offers no programmatic testing interfaces. 
\item \textbf{C2:} AVP's spatial user interface (UI) design creates fundamental challenges for UI exploration. The platform's 3D interaction model requires handling complex spatial relationships, motion tracking, and view management considering different head orientation of physical position,
which creates an expansive state space that necessitates efficient exploration methods.
\item \textbf{C3:} Detecting privacy violations in AVP network traffic requires a taxonomy that links low-level identifiers observed in packets to the high-level data categories used in privacy disclosures, \ycrv{but existing frameworks~\cite{trimananda2022ovrseen,zhan2024vpvet} are not sufficient to capture AVP-specific data types or maps traffic to Apple's multi-layered documentation system of policies, labels, and manifests}.
\end{packeditemize}

\bheading{\sysname.}
To tackle the challenges and bridge the gap, 
we propose \sysname, a novel testing framework for AVP applications. 
Our key insight is that effective AVP application testing requires coordinated control of both cyber (software) and physical (hardware) components. We term this approach \textbf{Coordinated Cyber-Physical Testing}, which enables comprehensive application analysis by bridging the gap between software interfaces (cyber-space) and hardware interactions (physical-space). 
To tackle C1,
we build a customized hardware automation system that uses Bluetooth Human Interface Device (HID) to generate user inputs in AVP, and implements a digital-to-physical pipeline for feedback-driven control.
To tackle C2,
we develop a  UI exploration engine for exploring the 3D UI states,
which incorporates specialized UI recognition and state pruning approaches.
To tackle C3,
we construct a systematic privacy analysis framework that captures privacy violations from network traffic based on a unified AVP privacy taxonomy.

We implemented a prototype of \sysname, and we evaluated it using a manually labeled ground-truth network traffic dataset consisting of 50 AVP applications.
\sysname can cover  151 out of 157 privacy violations found via manual exploration within 20 mins of auto-exploration.
\sysname exploration can yield 3.15$\times$ of traffic compared to the idle baseline.
In addition,
we performed a large-scale analysis on 324 
COTS AVP applications,
and we used \sysname to test each of them for 20 minutes and collect network traffic to study their privacy compliance.

\bheading{Findings.}
Our large-scale analysis (\autoref{sec:measure}) reveals widespread non-compliance with privacy disclosure requirements in the AVP app ecosystem. 188 of apps with observable network activity had one or more disclosure violations, and more than 60\% of the network traffic flows are not properly disclosed. 
The most common deficiencies stem from omissions and inaccuracies in privacy policies (\autoref{sec:policy}) and labels (\autoref{sec:label}). Similarly, privacy manifests, when present, are often incomplete or misrepresent actual data flows (\autoref{sec:manifest}).
A majority of non-disclosures are due to the use of SDKs and the inherent descriptions (\autoref{sec:sdk}), which are often inadequate. 
We further observe significant discrepancies between declared and actual purposes of data collection (\autoref{sec:purpose}), exacerbated by ambiguities in Apple's data taxonomy and insufficient tooling for SDK attribution, ultimately leading to developer confusion and inconsistent compliance (\autoref{sec:discussion:responses}).

\bheading{Contributions.}
We make the following contributions:

\begin{packeditemize}
    \item We introduce \sysname, the {\it first} end-to-end  automatic testing tool for detecting privacy non-compliance of COTS AVP applications. It consists of a customized AVP device control system, a UI explorer, and a privacy violation detector.
    \item We evaluate \sysname through a 50-app ground-truth dataset, and show that \sysname is effective.
    \item We use \sysname to perform a large-scale analysis on 324 AVP applications, with each app tested for 20 minutes, which reveal that widespread non-compliance issue. %
    \item We report our findings to the related parties, and we offer recommendations for addressing the issues.
\end{packeditemize}

\bheading{Responsible disclosure.}
We have reported all findings (privacy violations) to Apple and related app developers.
Their responses are discussed in~\autoref{sec:discussion:responses}.

\section{Background}
\label{sec:bg}

\subsection{Apple Vision Pro}
\label{sec:bg:avp}

\begin{figure}
    \centering
    \includegraphics[width=0.5\columnwidth]{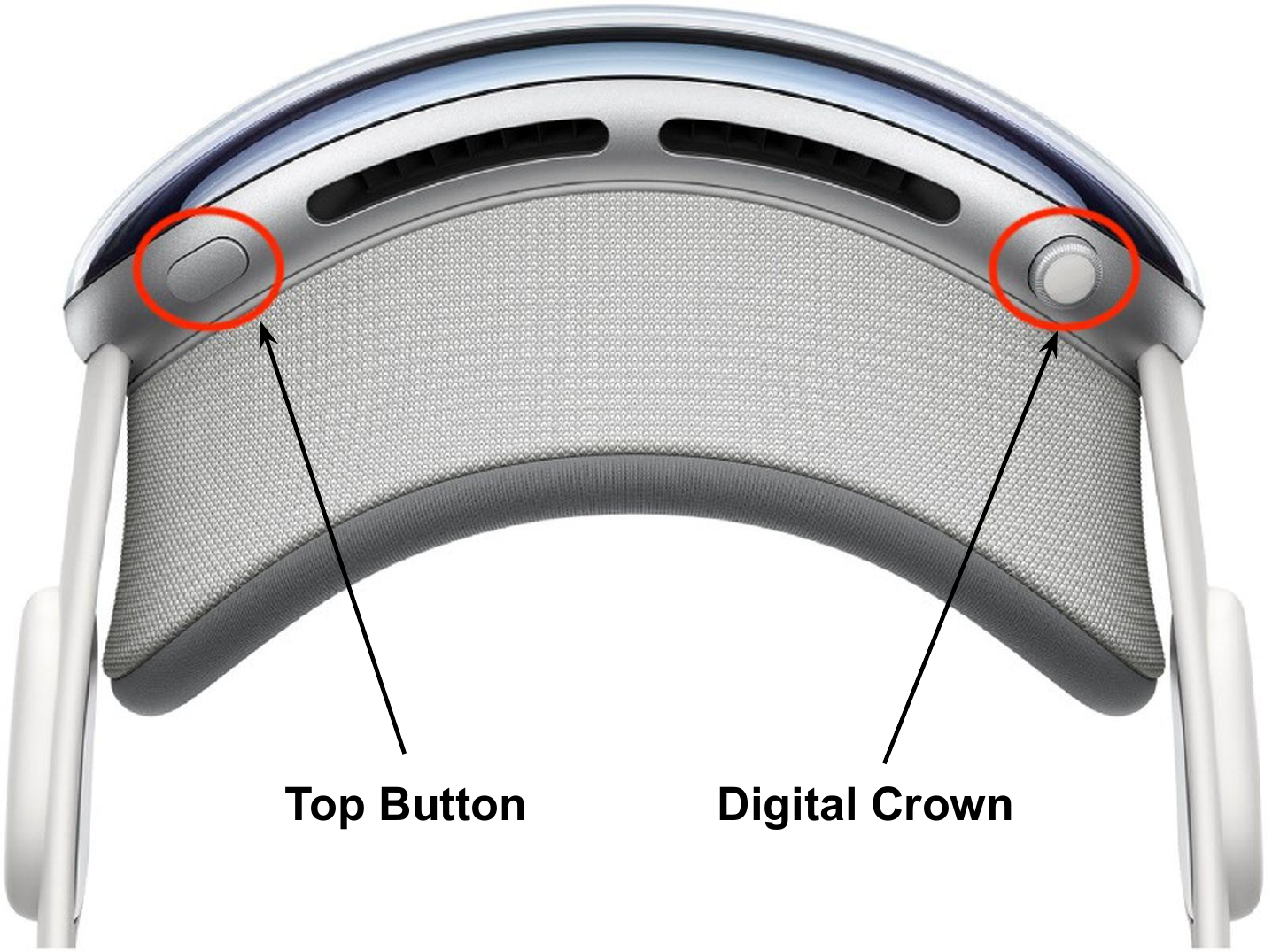}
    \caption{Apple Vision Pro}
        \label{fig:AVP}
\end{figure}

The Apple Vision Pro (AVP)~\cite{avp-web} represents a pivotal step in the evolution of spatial computing. Introduced in 2024 with a price tag of 3,499 USD, Vision Pro integrates dual high-performance chips to support real-time sensor fusion, low-latency environment mapping, and high-fidelity mixed-reality rendering. 
It runs a specialized Operating System called visionOS which was designed for XR devices.
Different from the popular Meta Quest series, AVP's control architecture relies on multimodal input combining precise eye tracking, hand gestures, and voice commands, which allows natural, controller-free interaction. Two hardware interfaces complement this input model (shown in~\autoref{fig:AVP}): the Digital Crown (right), which adjusts immersion levels and system view, and a Top button (left) for power, capture, and system resets.

\subsection{Privacy Documents in Apple Vision Pro}
\label{sec:bg:priv}

\bheading{Privacy policy.} A privacy policy is a legal document provided by app developers that outlines how an app collects, uses, and shares user data, as regulated by laws such as GDPR~\cite{GDPR} and CCPA~\cite{CCPA}. 
Each privacy statement in the privacy policy can be modeled as a tuple \texttt{(entity, action, data)}, 
describing what data is collected (or explicitly not collected) by whom~\cite{andow2019policylint, andow2020actions, cui2023poligraph}.
The \textit{entity} is typically categorized as a first party (the app developer) or a third party (e.g., advertising or analytics SDKs).
Without proper disclosure of these data behaviors in the privacy policy, the app poses privacy risks to end users and falls out of compliance with privacy laws.

\ignore{
Despite their importance, privacy policies are often lengthy and legalistic, making them difficult for users to understand and for auditors to compare with actual data flows.
To address this, researchers have developed both symbolic NLP techniques~\cite{andow2019policylint, andow2020actions, cui2023poligraph} (e.g., part-of-speech tagging~\cite{martinez2012part}) and statistical methods~\cite{zimmeck2017automated} (e.g., machine learning) to automate privacy policy analysis.  
However, symbolic methods rely on rigid rules usually extracted from mobile application domain and often fail to generalize across different contexts (e.g., IoT, Vision Pro)~\cite{qin2025automated}, while statistical approaches require extensive manual annotation~\cite{zimmeck2017automated,wilson2018analyzing, bui2021automated, harkous2018polisis}.
Recent advances in Large Language Models (LLMs) have substantially improved automated privacy policy analysis. 
LLMs can extract relevant information from complex and diverse policy documents without manual annotation, and studies~\cite{goknil2024privacy, tang2023policygpt, chen2025using, rodriguez2024large} show they generalize better across different formats and domains than traditional methods.}

\bheading{Privacy label.}
To improve transparency and user awareness, Apple requires developers to complete privacy labels for all Vision Pro apps submitted to the App Store~\cite{visionpro-privacy-label}. These labels provide standardized, easily readable summaries of the types of data an app may collect and how that data may be used or shared~\cite{Apple-privacy-label}.
The privacy disclosure of the
privacy label follows a four-layer taxonomy: data usage, purpose, data type and data item.
A key distinction of Vision Pro privacy labels is the inclusion of unique data categories that reflect the device’s advanced spatial and biometric sensing capabilities. 
Unlike mobile apps, Vision Pro applications may collect detailed environmental and interaction data, such as room mapping, scene mesh information, and user biometric signals like hand and eye movement data~\cite{Apple-privacy-label}.
These expanded categories go beyond traditional location or usage data, underscoring the need for greater transparency in how XR technologies collect and process sensitive information, a focus that our study directly addresses.

\bheading{Privacy manifest.}
The privacy manifest is a machine-readable declaration (typically named \cc{PrivacyInfo.xcprivacy}) that developers are required to include with the app binary when submitting to Apple’s App Store, as mandated since May 2024~\cite{Privacy-manifests-specification}.
Unlike privacy policies and privacy labels (which are written for end users), the privacy manifest primarily discloses data collection behaviors to the App Store and automated compliance tools. Its main purpose is to facilitate privacy compliance checks during the app review process.
An inaccurate privacy manifest can mislead Apple’s review process, resulting in incorrect verification of privacy disclosures and inconsistencies with the information reported in privacy labels. 
Therefore, our study also evaluates non-compliance in privacy manifests.
\looseness=-1

\section{Overview}
\label{sec:overview}

\bheading{Goal.}
We aim to systematically evaluate privacy compliance in AVP applications through a dynamic analysis pipeline. 
Our objective is to develop an automated testing framework that triggers and captures network communications from commercial off-the-shelf (COTS) AVP applications \textit{without requiring root access or application source code}. 
The framework analyzes
privacy violations by identifying discrepancies between observed network traffic and the privacy declarations made by applications. We define a privacy compliance violation as any instance where privacy-sensitive data is transmitted to external parties without proper disclosure in the application's privacy documentation. Our analysis considers three key privacy documentation sources associated with AVP applications: privacy policies, privacy labels, and privacy manifests (\autoref{sec:bg:priv}).

\subsection{Challenges}
We face three main challenges when attempting to automatically test AVP applications to capture network traffic.

\bheading{Challenge 1: Lack of programmatic interfaces for input simulation and device control.}
The first challenge stems from AVP's restrictive system architecture that prevents programmatic input simulation, which is a fundamental requirement for automated testing. Unlike iOS devices where jailbreaking~\cite{ios-jailbreak1} enables root access, AVP follows a strictly closed system model similar to other commercial XR platforms. This architectural choice precludes the use of established iOS testing frameworks like Macaca~\cite{macaca} or XCUITest~\cite{XCUItest} that require root access. Furthermore, while Android devices support debugging interfaces like ADB~\cite{ADB} even without root access, AVP provides no equivalent programmatic control interface. The platform also lacks standard input/output ports, preventing direct device control through conventional peripherals such as keyboard and mouse. 
These limitations create significant barriers for implementing automated testing procedures.

\begin{figure}[t]
\begin{minipage}[t]{0.49\columnwidth}
\centering
 \includegraphics[width=\linewidth]{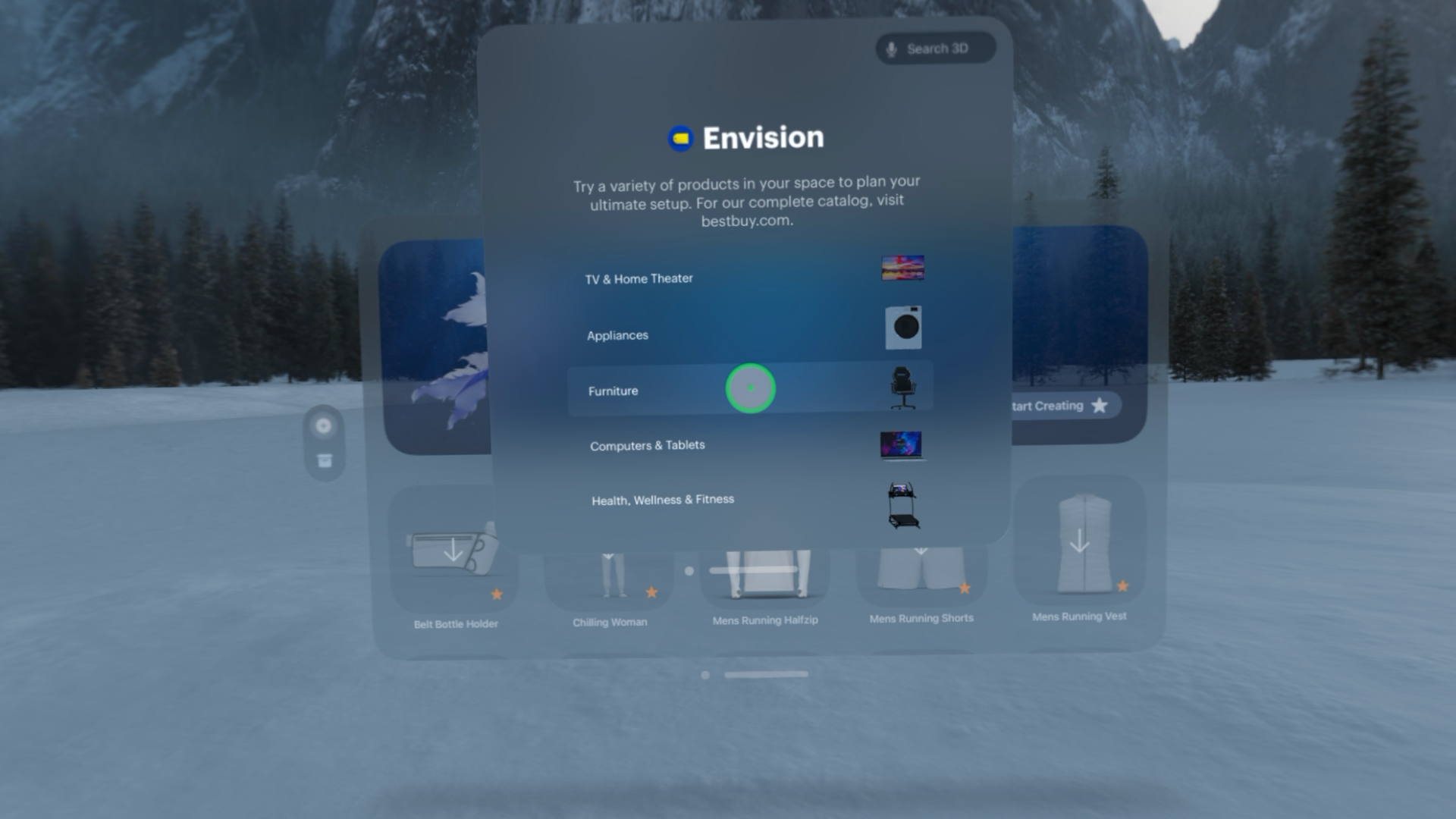}
\end{minipage}
\hfill %
\begin{minipage}[t]{0.49\columnwidth}
\includegraphics[width=\linewidth]{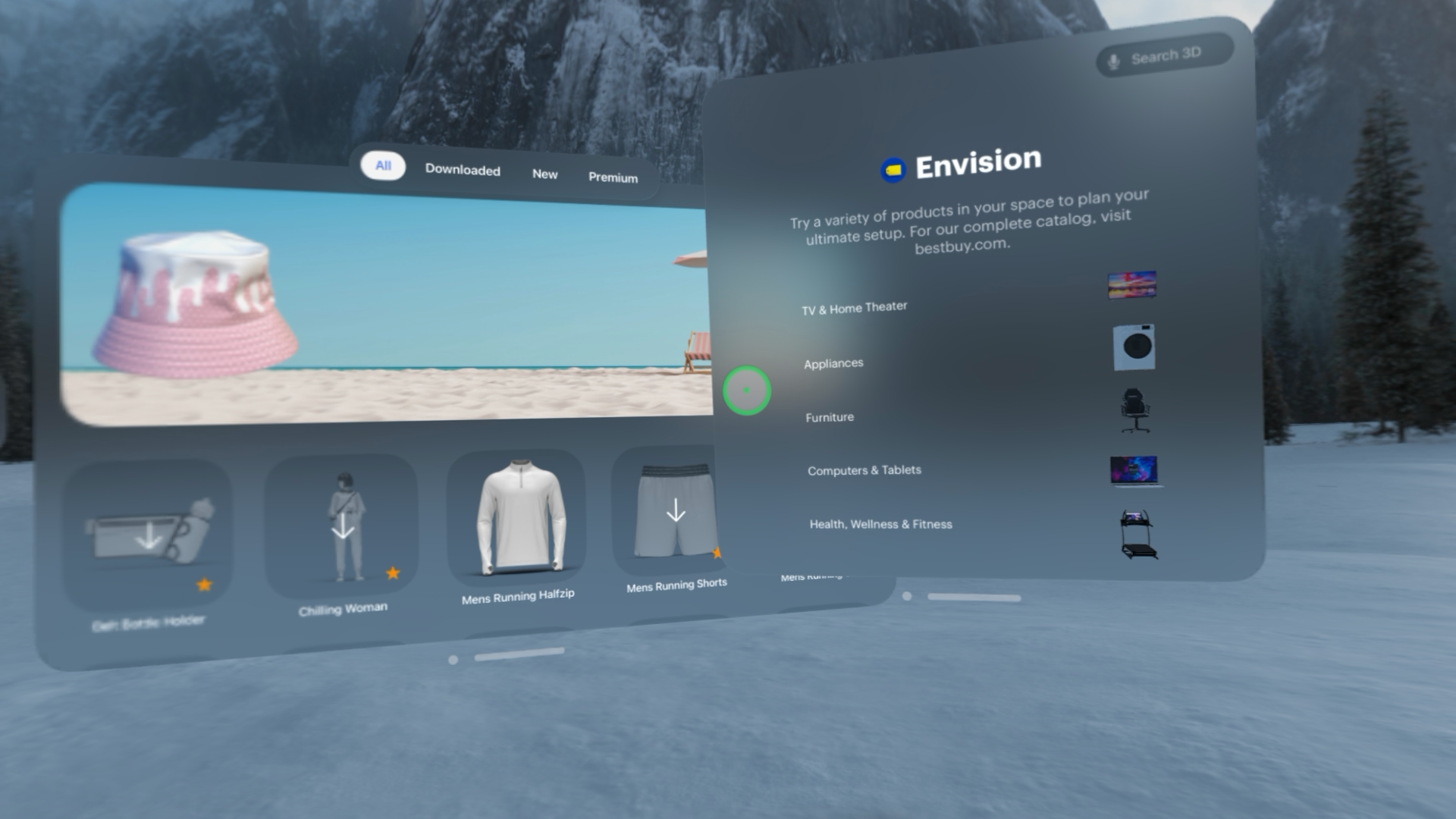}
\end{minipage}
\caption{Same UI with different depth and orientation.}
\label{fig:ui-depth}
\end{figure}

\bheading{Challenge 2: Difficulty in exploring 3D UIs.}
Different from Android-based applications,
AVP applications do not have XML-like UI documents.
In addition,
AVP applications present unique challenges for automated UI exploration due to their three-dimensional nature. Unlike traditional 2D interfaces where UI states are deterministic, 3D interfaces in AVP exhibit state variability based on user viewpoint parameters such as head orientation and physical position. This variability complicates state identification and deduplication since identical functional states may appear visually different (\autoref{fig:ui-depth}). 
This exponential growth in the state space due to duplicated states makes exhaustive exploration computationally intractable without intelligent pruning strategies.

\bheading{Challenge 3: \ycrv{Insufficient} privacy taxonomy for AVP traffic.}
Detecting privacy violations in AVP traffic requires a comprehensive taxonomy that maps lower-level terms (e.g., \textit{``hand dimensions''} ) in privacy policy to upper-level privacy data (e.g., \textit{``vr movement''}) in privacy documentation declarations. 
While prior work has developed privacy taxonomies for other XR platforms (\eg, VPVet~\cite{zhan2024vpvet} and OVRSeen~\cite{trimananda2022ovrseen}), \ycrv{these frameworks do not capture AVP-specific data categories or the relationships between network traffic and AVP's multi-layered privacy documentation system comprising policies, labels, and manifests due to ambiguity in those Apple specific privacy documentations.
}

\subsection{Overview of \sysname}
To systematically analyze privacy compliance in AVP applications via examining network traffic, we propose \sysname, a novel testing framework that addresses the aforementioned challenges. Our key insight is that effective AVP application testing requires coordinated control of both cyber (software) and physical (hardware) components. We term this approach \textbf{Coordinated Cyber-Physical Testing}, which enables comprehensive automatic testing by bridging the gap between software interfaces (cyber-space) and hardware interactions (physical-space). Based on this insight, we design \sysname with three integrated components: \compone for automated device control, \comptwo for AVP UI exploration, and \compthree for privacy violation detection. Together, these components form an end-to-end system that automates privacy compliance testing of AVP applications.

\bheading{\compone: Hardware-based input simulation (addressing Challenge 1).}
We develop a hardware-based automation system that circumvents AVP's restrictions through a combination of techniques. At its core, the system uses a programmable Bluetooth Human Interface Device (HID) to generate virtual cursor inputs in AVP. We augment this with 
physical button control mechanisms. 
This comprehensive approach enables programmatic control of the AVP device without requiring system modifications or root access.

\bheading{\comptwo: UI exploration for AVP applications (addressing Challenge 2).}
We develop a specialized UI exploration engine that efficiently handles AVP's 3D UIs. Our approach first deploys screen mirroring for visual feedback processing, then projects 3D scenes onto 2D planes for simplified analysis. 
In addition, we design a precise cursor movement algorithm to control \compone to reach our target coordinates based on screenshots.
We employ machine learning-based UI recognition to identify equivalent interface states across different exploration runs, accounting for visual variations caused by 3D perspective changes. 
We construct a state transition graph based on interactive UI elements, enabling efficient pruning of redundant states.

\bheading{\compthree: Privacy violation detection from network traffic (addressing Challenge 3).}
We develop a systematic privacy-violation detection module that identifies six types of privacy non-compliance by cross-checking data terms in network traffic against three key privacy documents.
To achieve this, we extend prior taxonomies with AVP-specific privacy categories and data types derived through automated keyword extraction from our evaluation dataset. Our framework establishes unified taxonomies that align network traffic patterns with privacy documentation sources including manifests, labels, and policies. Using these aligned taxonomies, we implement a violation detection model that systematically identifies discrepancies between observed runtime network behavior and declared privacy practices.

\section{Design of \sysname}
\label{sec:sys}

\begin{figure}[t]
    \centering
    \includegraphics[width=.8\linewidth]{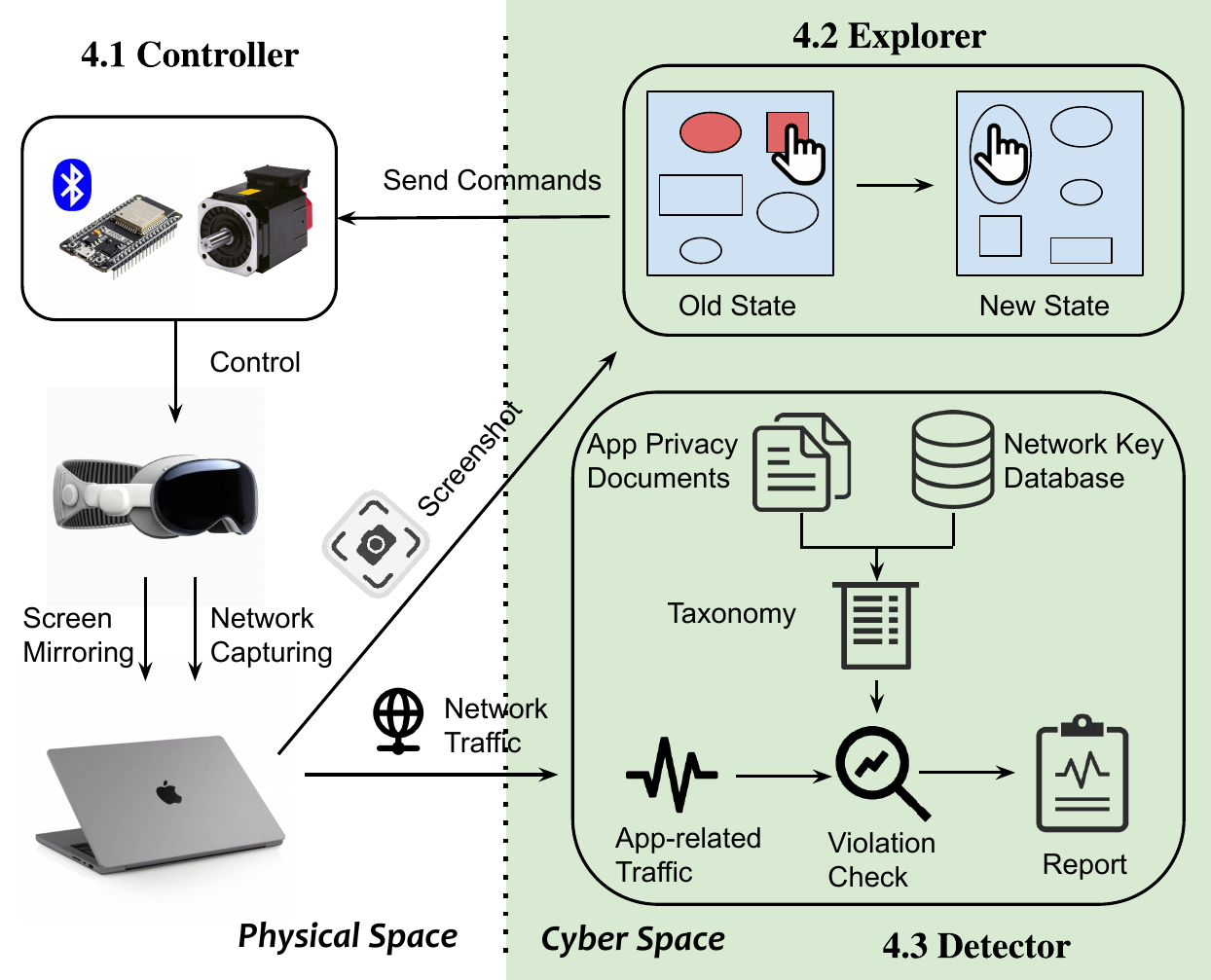}
    \caption{Overview of \sysname.}
    \label{fig:system_workflow}
\end{figure}

\autoref{fig:system_workflow} illustrates the workflow of \sysname.
During the setup process, \sysname utilizes screen mirroring to replicate the screen of the AVP device onto a Macbook.
The Macbook periodically captures screenshots,
which are sent to \comptwo for analysis.
Given a list of applications,
\comptwo will first instruct \compone to open one target application,
then start the exploration through UI analysis.
During the exploration,
\comptwo will communicate with \compone frequently to understand the UI contents to decide the next steps.
During the exploration,
the network traffic will be captured and sent to \compthree, along with the application's privacy documents,
for checking privacy violations.
In the end, a report will be created for the target application, detailing the traffic that includes privacy violations and specifying the type of data involved in these violations.

\subsection{\compone: Simulating User Interactions}
\label{sec:sys:compone}

Although Apple provides the AVP developer strap~\cite{AVP-strap}, its functionality is limited to screen capturing, firmware downgrading, and fast file transfer, and it lacks the ability for programmatic input.
In this subsection,
we detail our design to overcome the lack of programming interfaces for input simulation.
In particular, 
we aim to tackle two important problems:
i) simulating user inputs (e.g., move cursors in the XR scene, touch virtual objects, etc);
ii) pressing physical buttons on AVP device for XR-specific control that cannot be simulated.

\bheading{i) Simulating User Inputs.}
\label{sec:sys:compone:input}
Emulating an input device (\eg, mouse) purely in software is not feasible on AVP due to several reasons. 
First, AVP does not expose synthetic input APIs for third-party to call.
Moreover, AVP's Human Interface Device (HID) stacks (including both USB and Bluetooth devices) are handled at the system level: only external accessories that i) present as standards-compliant HID devices (e.g., BLE HOGP or Bluetooth Classic HID) and ii) complete secure pairing/bonding are accepted.
As a result,
software-based ``virtual'' HID devices created by third-party applications are not recognized by AVP, and cannot be paired or used as inputs.

As software-based simulation cannot work,
we resort to hardware-based simulation.
We first attempted to use laptops (e.g., Macbook) to simulate as HIDs.
However,
general-purpose computers typically operate as HID hosts (Central/BR/EDR host) and do not present a valid HID peripheral without dedicated firmware. 
Next,
we have tried to use Bluetooth mouses such as Logitech mouses.
While such devices can indeed pair with AVP,
they do not provide programming interfaces for automatic control,
as they need to be manually operated. 
To tackle this issue,
we design a customized HID using an ESP32 device that presents a compliant Bluetooth HID keyboard and mouse with bonding persistence,
following the Apple accessory guidelines~\cite{Apple-accessory}.
The ESP32 device will listen on the commands from serial port, then emulated mouse and keyboard control event (mouse movement, click, keystroke, etc) via Bluetooth.

\begin{figure}[t]
    \centering
    \includegraphics[width=0.77\columnwidth]{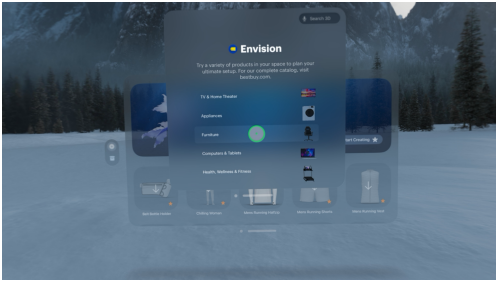}
    \caption{Overlapping UIs from multiple applications.}
        \label{fig:ui-overlap}
\end{figure}

\bheading{ii) Pressing Physical Buttons.}
With the simulated user inputs,
we can already have the keyboard/mouse input to the AVP.
However, AVP still limit the ability of external keyboard/mouse input;
there are certain scenarios where physical pressing the hardware buttons is necessary.
For example,
in AVP, when the user's head moves, the view will not be automatically centered.
The user needs to press and hold the Digital Crown button (\autoref{fig:AVP}) to center the view.
This is very important,
as the  UI difference will affect the exploration in \comptwo.
Moreover,
AVP does not provide simple ways for killing an application.
This is also critical for us,
as UIs of different applications may overlap and confuse \comptwo.
One example is shown in~\autoref{fig:ui-overlap}.
To do so,
it requires the user to press Digital Crown and Top buttons  together to call out the \textit{Force Quit} window (similar to Windows Task Manager)~\cite{AVP-kill-app}, 
then choose the app to kill.
Therefore, pressing the physical buttons is required for automatically exploring AVP applications.
To achieve the goal, we need to be able to apply proper forces to press the physical buttons as if the user is pressing them.
To do so, 
we use servo motors~\cite{servomotor} to provide the forces.
To make sure that
the servo motors can be programatically controlled,
we integrate it with the user input simulation system in~\autoref{sec:sys:compone:input}.
Therefore,
we can directly control the motors for pressing the physical buttons through the same ESP32 interface.
\looseness=-1

\subsection{\comptwo: Exploring AVP UI States}
\label{sec:sys:comptwo}

To perform exploration of AVP UIs,
\comptwo is designed with three major sub-components:
i) \textit{screen capturer} for taking screenshots of AVP UI;
ii) \textit{virtual cursor handler} for handling the movements of virtual cursor using \compone;
iii) \textit{state explorer} for exploring the UI states using the virtual cursor controller.

\bheading{i) Screen Capturer.}
\label{sec:sys:comptwo:screen}
The first step is to capture the screenshot from AVP.
While the AVP device provides the screenshot functionality,
it is inconvenient to capture the screenshot programatically and send the image out for processing.
To automate the screen capturing process,
we use the AVP's built-in AirPlay functionality to mirror the AVP screen to a MacBook Pro,
then perform recording and screenshot on the MacBook Pro.
After capturing the screensots,
we will perform analysis on the screenshots, which will inform us next steps (e.g., move cursor, click button, etc).

\bheading{ii) Virtual Cursor Handler.}
\label{sec:sys:comptwo:cursor}
We design a calibrated cursor positioning and moving system that maps 2D screen coordinates (from the mirrored screenshots) to 3D space interactions (in the AVP device screen) to precisely control the cursor in the 3D space in AVP apps.

\iheading{Locating the cursor from 2D screenshots.}
Unlike traditional smartphone UI exploration systems that can query cursor positions directly,
AVP does not provide ground truth cursor coordinates.
We must therefore locate the virtual cursor from screenshots.
The default cursor appears as a small white dot that is difficult to distinguish from application UI elements and backgrounds.
To address this challenge,
we modify the cursor appearance in system settings by adding an outer circle and changing its color to green.
We then apply two computer vision techniques to detect the cursor:
(1) color filtering to isolate green pixels, and
(2) Hough Circle Transform ~\cite{Circle_Hough_Transform} to identify the coordinates of the green circle with known radius.
\looseness=-1

\iheading{Controlling the cursor movement in the 3D space.}
Since AVP does not provide absolute position control for cursor movement,
we design an algorithm to move the virtual cursor to target positions.
The key challenge is that 2D screenshots do not accurately represent depth in 3D space.
Objects at different depths appear at similar 2D coordinates but require different movement distances to reach.
Therefore, we cannot directly map 2D screen distances to cursor movements in 3D space.
We address this challenge using a binary search approach.
Given the current cursor position $(x,y)$ and target position $(x',y')$ in the screenshot,
we move along each axis independently: first along the X-axis to reach $(x',y)$, then along the Y-axis to reach $(x',y')$.
To move from $x$ to $x'$, we perform the following steps:
1) Move the cursor toward $x'$ by $\lambda$ units (a large initial step size) in \compone,
and repeat until the cursor moves past $x'$.
2) Let $x_1$ denote the new cursor position. If $|x_1-x'| \leq \delta$ (where $\delta$ is a small threshold), the cursor has reached the target. Otherwise, proceed to step 3.
3) Set $\lambda = \lambda/2$ and $x=x_1$, then return to step 1.
\looseness=-1

\bheading{iii) State Explorer.}
\label{sec:sys:comptwo:explorer}
To systematically explore the UI states in AVP, 
we design a state explorer that constructs the UI state graph of the target  application on-the-fly, as there is no existing interface to obtain such information.

\iheading{Definitions.} 
To comprehensively explore the AVP application, we use a \emph{state graph} to represent the application's state transition. 
The state graph is a directed graph, where the nodes are the UI states of the application, and the edges are the transitions between states.

\begin{figure}
    \centering
    \includegraphics[width=0.77\linewidth]{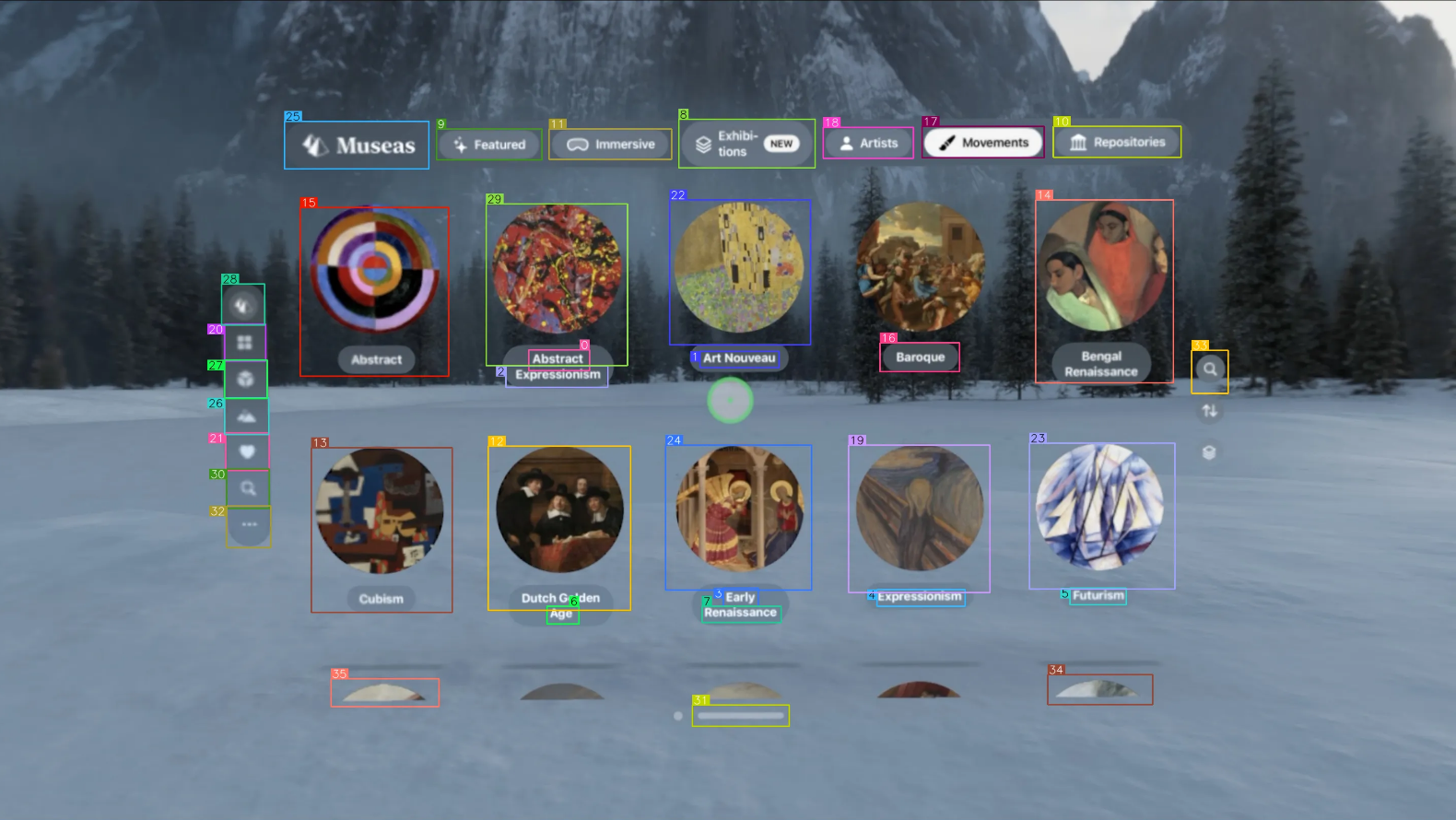}
    \caption{UI element detection.}
        \label{fig:ui-element}
\end{figure}

\begin{packeditemize}
    \item \textit{Node (State).}
The node (\ie, state) of the state graph is defined as the current UI in the 2D screenshot,
which consists of buttons, images, texts, \etc.
To obtain a comprehensive list of objects in the 2D screenshot to represent the state,
we use an object detection model pre-trained for UI elements to precess the 2D screenshot. 
The model will output the detected UI elements with detailed information, such as their coordinates and the text contents,
which is recorded for denoting the state.
One example is shown in~\autoref{fig:ui-element}.

\item \textit{Edge (State transition).}
An edge in the state graph represents a transition between two states, triggered by a UI element.
In our implementation, edges correspond to clickable UI elements (buttons) that cause state transitions.
An edge can be a forward edge, representing entering a new state, or a backward edge, representing a return to a previous state.
\looseness=-1
\end{packeditemize}

\iheading{Exploration algorithm.}
\autoref{alg:app-exploration} (\autoref{app:algorithm}) presents our state exploration algorithm.
The algorithm maintains a state graph $G$ that records all discovered states (nodes) and their transitions (edges).
Starting from the initial state of the application (\cc{Main()}),
the algorithm performs a depth-first search (DFS) to systematically explore all reachable states.
For each state, the algorithm identifies all clickable UI elements (buttons) and attempts to click each unexplored button.
After each click, it captures a screenshot and extracts UI elements to determine the current state (\cc{CheckCurrentState()}).
If the current state is new, it is added to the state graph along with the transition edge.
If the state has been visited before, the algorithm resumes exploration from that state by clicking its remaining unexplored buttons (\cc{ResumeExploreState()}).
When all buttons in a state have been explored, the algorithm attempts to return to the previous state by clicking the back button (\cc{TryClickBackButton()}).
If no unexplored buttons remain in the current exploration path, the algorithm selects another state with unexplored buttons from the state graph, jumps to that state by restarting the application and replaying the path from the initial state (\cc{JumpToCurrentState()}), and continues exploration (\cc{ResumeExploreState()}).
This process continues until all states have been fully explored.

\iheading{Navigating back to previous states.}
Our DFS-based exploration requires a mechanism to return to previously visited states.
In AVP, most applications provide a \textit{back} button at the top left corner of the window, similar to Android.
Clicking this button returns the application to the previous state.
However, unlike Android where the back button is system-level, AVP applications do not always include this button.
Specifically, pop-up windows often lack a back button.
In such cases, after exploring all states within the pop-up, we must close the window to return to the previous state.
We achieve this by using a cursor movement pattern to click the \textit{Close} button~\cite{AVP-move-close} located in the bottom window bar.

\iheading{Returning to the initial state.}
The exploration algorithm requires a method to return to the initial application state when jumping between unexplored states (\cc{JumpToCurrentState()} of \autoref{alg:app-exploration} in \autoref{app:algorithm}).
In AVP, simply closing all windows does not reset the application.
When reopened, the application resumes at its last state rather than the initial state.
To reset the application, we use the \textit{Force Quit} feature.
We first trigger \textit{Force Quit} through \compone, then execute a cursor movement pattern to terminate the application, ensuring it restarts from the initial state.

\iheading{Same-state pruning.}
Unlike 2D scenarios (e.g., Android) where identical states produce identical UI layouts, same AVP states can have small differences in UI layouts with user viewpoint parameters such as head orientation and physical position.
To handle these minor UI variations,
we compute state similarity using the (position, size) of the detected UI elements.
If the similarity exceeds a threshold, we consider the two states identical.
During exploration, we maintain all discovered states in the state graph.
When transitioning from state A to state B,
we check whether state B matches any adjacent explored state of A.
If a match is found, we return to state A.
Otherwise, we add state B as a new adjacent state of A and continue exploration.
This pruning strategy significantly reduces the number of states explored.

\subsection{\compthree: Detecting Privacy Violations}
\label{sec:sys:compthree}

\compthree analyzes network traffic captured during app exploration to detect privacy violations by comparing transmitted data against privacy disclosures.
It consists of three steps: (1) building a unified privacy taxonomy that maps data types from privacy policies, labels, manifests, and network traffic to a common representation; (2) extracting privacy-related keys from network traffic and mapping them to taxonomy nodes; and (3) identifying violations by comparing detected data types against privacy disclosures.

\bheading{Privacy Taxonomy.}
To detect privacy violations, we build a unified taxonomy that maps data types from multiple sources---privacy policies, privacy labels, privacy manifests, and network traffic---to a common representation.
We construct our taxonomy by extending VPVet's taxonomy~\cite{zhan2024vpvet} with AVP-specific data types through a four-step process. 
First, we extract data types from privacy documents. We use the \cc{PrivBERT} model
to extract privacy statement triplets \cc{⟨entity, action, data type⟩} from privacy policies, directly use Apple's predefined data type categories~\cite{Apple-privacy-label} from privacy labels, and parse the \cc{NSPrivacyCollectedDataTypes} field from privacy manifests.
Second, we map all extracted data phrases to VPVet's existing taxonomy using synonym matching and collect phrases that fail to match as \textit{unmapped phrases}, which represent AVP-specific data types not covered by traditional mobile privacy taxonomies.
Third, we discover new data types from unmapped phrases using the embedding method in VPVet (see~\autoref{alg:embedding-discovery} in~\autoref{app:algorithm}): we manually select a representative phrase as a new data type candidate, compute similarity between the candidate and remaining phrases, group similar phrases (similarity $\geq 0.8$) under the same data type, and repeat until no new types can be identified.
Fourth, we manually insert newly discovered data types into appropriate positions in the taxonomy hierarchy (e.g., biometric-related types under the \cc{PII} branch), with the grouped phrases from the previous step becoming synonyms for each new data type to enable phrase-to-term mapping.

\ycrv{Concretely, we pair each lower-level data type with its upper-level node
by annotating is-a (hypernym) relations: a lower-level type $B$ is linked to an
upper-level node $A$ when $B$ \emph{is an} $A$. For example, the traffic-side
type \cc{user id} is paired with the Apple privacy-label category \emph{User
ID}. Two annotators performed this labelling independently and a third
adjudicated the disagreements.}
Starting from VPVet's 106 data type (after filtering 1 Android-specific data type), 
we extend our taxonomy to include new data type.
We add 12 data types from privacy policies and 35 data types from privacy labels.
Finally, we have a total of 153 data type in our taxonomy.

\bheading{Network Traffic Key Extraction.}
To identify privacy-related data in network traffic, we extract all key-value pairs from HTTP requests by parsing three sources: (1) HTTP headers, (2) URL query parameters, and (3) request bodies (including nested JSON/XML structures). For nested data structures, we recursively flatten them to obtain all key-value pairs. We then map extracted keys to taxonomy nodes using three matching strategies. First, \textit{synonym matching} maintains a synonym list for each taxonomy node and maps keys that exactly match any synonym (e.g., \cc{"user\_id"}, \cc{"uid"}, and \cc{"userId"} all map to \cc{user identifier}). Second, \textit{key pattern matching} uses regular expressions to match common naming patterns (e.g., keys ending with \cc{"\_id"} or \cc{"\_token"} map to identifier-related nodes). Third, \textit{value pattern matching} uses regular expressions to match value formats for certain data types (e.g., values matching \cc{xxx.xxx.xxx.xxx} map to \cc{IP address}).

\bheading{Privacy Violation Detection.}
We compare data types found in network traffic against three privacy disclosure sources to identify violations.
A unique violation is defined as a distinct \cc{(key, category)} pair that appears in network traffic but is not covered by the app's privacy documentation:
 1) a \textit{policy violation} occurs when the data type is not disclosed in the privacy policy;
2) a \textit{label violation} occurs when it is not declared in the App Store privacy label; 
and 3) a \textit{manifest violation} occurs when it is not listed in the privacy manifest.
For example, if an app transmits \cc{x-device-id} (mapped to ``Identifiers'') but does not declare ``Identifiers'' in its privacy documents, this constitutes a violation.

\ycrv{More details about  the three matching mechanisms, first-party
versus third-party classficiation, and why neither the policy nor the
label can serve as ground truth
are presented in \autoref{app:detector}.}

    \section{Evaluation}
    \label{sec:eval}

\begin{figure}
    \centering
    \includegraphics[width=0.7\linewidth]{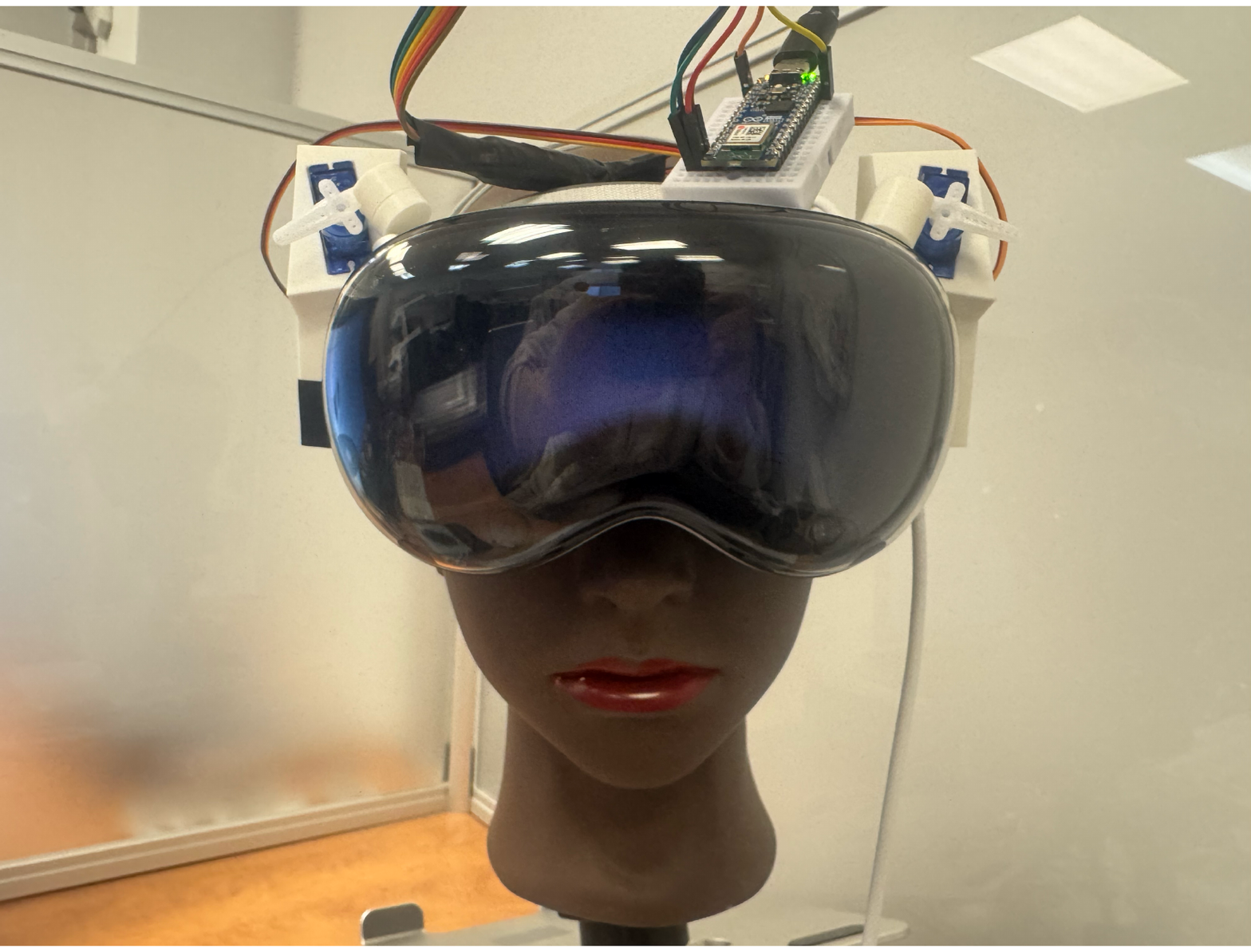}
    \caption{Implementation of \compone.}
    \label{fig:controller}
\end{figure}

\bheading{Implementation.}
We implement a prototype of \sysname, 
which consists of 286 lines of ESP32 code 
and about 3,000 lines of Python code.
\autoref{fig:controller} shows our implementation of \compone.
We stabilize the AVP device using a makeup training head mounted on a clamp stand during experiments.
We implement the customized HID using an Arduino ESP32-S3 microcontroller with built-in Bluetooth Low Energy support. 
To construct UI states from screenshots, we deploy a pretrained OmniParserV2 model~\cite{omniparserv2}. 
We capture network traffic from the AVP device using a Man-in-the-Middle (MITM)  approach.
More details are presented in~\autoref{sec:impl}.
    
    \bheading{Experimental Setup.}
    In our experiment, the AVP device is running visionOS 2.1.
    The AVP screen is mirrored to a MacBook M2 Pro with 32GB RAM running macOS Sequoia. 
    The same MacBook runs \cc{mitmproxy} to capture network traffic from the AVP.
    We run the \cc{OmniParserV2} model on a Lambda Vector Pro server with Threadripper PRO 7985WX CPUs (64 cores), two NVIDIA RTX A6000 GPUs (48GB each), 256GB RAM, and 10TB NVMe SSD. 
    Screenshots captured on the MacBook are sent to the Lambda server for processing.
    Both machines are connected to the same local network to minimize communication latency.

    \bheading{Evaluation Process.}
    We created an automated script to download and install target apps from the AVP App Store using \sysname based on App IDs from our dataset.
    For each app in the list,
    we install it on AVP and open it by searching for the application name in the \cc{Spotlight} search.
    We then use the Digital Crown button to center the view and begin testing for 20 minutes.
    After testing completes, we force quit the application and proceed to the next one.
    For each app, we used screen recording to record all explorations when \sysname is exploring the app. We also record the network traffic triggered during exploration.
    We use the timestamps of starting/ending each app to segment the network traffic for each app.
    \looseness=-1

    \subsection{Datasets}
    \label{sec:eval:data}

We constructed our AVP app dataset by crawling the 3rd party website \cite{vision-dir} (the official VisionPro App Store does not provide app list) that list all available AVP apps, which contained 671 apps, of which 460 were free. For the free apps, we excluded 86 apps that were not applicable to our analysis: 16 Apple system apps (e.g., Safari) that are out of scope, 4 Microsoft apps (e.g., Word) whose IPA files could not be acquired through standard methods, 20 Apple Arcade games that require an active subscription despite being listed as free, and 46 apps requiring visionOS versions newer than our test device (visionOS 2.1), including 7 apps requiring visionOS 2.2-2.6 and 39 apps requiring visionOS 26.x. After applying these filters, our final dataset comprises 374 AVP apps that we successfully tested and analyzed.
\ycrv{We used the third-party directory only to enumerate which titles are
available for visionOS, as the official App Store exposes no such listing; every
app we analyze was downloaded from the official App Store through \cc{ipatool},
and privacy labels were collected from App Store.
The number of apps per category is proportional to that category's size,
and within each category the apps are selected at random.}

    \bheading{Ground Truth Dataset.}
    To evaluate \sysname's ability to detect privacy violations, we construct a ground truth dataset through manual exploration of AVP apps in our dataset.
    We selected 50 apps spanning top-10 popular categories. %
    As the number of available apps varies across categories, we selected apps proportionally based on the size of each top-10 category, and each category is represented by at least two apps in our dataset.
    During manual exploration, we trigger all possible interactive UI elements (e.g., buttons and objects) in the app to cover as many states and network requests as possible. The exploration stops when no interactive elements remain. In cases where there are multiple pages of the same type of items (e.g., a BestBuy store page), we only explore one page, as interacting with different items typically generates the same type of network traffic.
    We manually analyze the network traffic to identify unique network requests, extract keys, and check for violations. We consider network requests that retrieve or send the data type keys (regardless of the value) as identical.
    If any network key maps to our privacy taxonomy but not declared in the privacy documentation, we mark it as a potential violation.
    Manual exploration and analysis required approximately 48 person-hours for the 50 apps.
    This dataset is used for our evaluation~(\autoref{sec:eval}).

    \bheading{Large-scale AVP App Dataset.}
    We use the remaining 324 apps in our dataset for large-scale analysis.
    For each app, we extracted metadata and privacy information (privacy policy URLs and privacy labels) from the AVP App Store. 
    We also downloaded the IPA files using \cc{ipatool} and extracted the privacy manifests.  
    We use this dataset for large-scale analysis (\autoref{sec:measure}) and to build our network key dataset, which we map to our privacy taxonomy as described in~\autoref{sec:sys:compthree}.

\begin{table}[t]
\centering
\caption{Network traffic and privacy violation detection results for 31 apps. \textbf{GT}: Number of unique network requests found during manual exploration; \textbf{Auto}: Number of unique network requests found by \sysname; \textbf{GT Vio}: Number of unique violations found during manual exploration; \textbf{Auto Vio}: Number of violations found by \sysname. Some violation numbers exceed network request counts because a single request may contain multiple violations.}
\label{tab:combined-eval}
\scriptsize
\setlength{\tabcolsep}{4pt}
\begin{tabular}{l l r r r r}
\toprule
\textbf{Category} & \textbf{App} & \textbf{GT} & \textbf{Auto} & \textbf{GT Vio} & \textbf{Auto Vio} \\
\midrule
\multirow{6}{*}{Productivity}
 & Rad Timer & 2 & 1 & 0 & 0 \\
 & FreelanceKit & 4 & 4 & 0 & 0 \\
 & Cardhop & 5 & 1 & 13 & 13 \\
 & Focus - Timer & 28 & 12 & 0 & 0 \\
 & FocusBeats & 4 & 3 & 0 & 0 \\
 & Flippy & 7 & 3 & 12 & 11 \\
\midrule
\multirow{4}{*}{Utilities}
 & Craft & 67 & 10 & 0 & 0 \\
 & Broadcasts & 312 & 238 & 0 & 0 \\
 & Qlone & 2 & 1 & 0 & 0 \\
 & WWidgets & 17 & 6 & 4 & 2 \\
\midrule
\multirow{2}{*}{Games}
 & Retrogram & 26 & 9 & 21 & 21 \\
 & Things & 78 & 40 & 0 & 0 \\
\midrule
\multirow{4}{*}{Entertainment}
 & Callsheet & 413 & 374 & 0 & 0 \\
 & Turn Off the Lights & 21 & 5 & 0 & 0 \\
 & Wet Your Beak & 45 & 24 & 2 & 2 \\
 & Paramount+ & 150 & 60 & 8 & 8 \\
\midrule
\multirow{5}{*}{Lifestyle \& Health}
 & alo Sanctuary & 130 & 102 & 4 & 3 \\
 & Decathlon USA & 21 & 8 & 4 & 4 \\
 & Mindr & 5 & 1 & 0 & 0 \\
 & Bible & 15 & 15 & 14 & 13 \\
 & Best Buy & 44 & 44 & 12 & 12 \\
\midrule
\multirow{4}{*}{Education}
 & Chemistry & 8 & 8 & 9 & 9 \\
 & Drawing Desk & 7 & 7 & 0 & 0 \\
 & Foxar & 81 & 81 & 4 & 4 \\
 & Inviewer & 9 & 3 & 24 & 23 \\
\midrule
\multirow{5}{*}{Creative}
 & Kineo & 1 & 1 & 5 & 5 \\
 & Spatial Station & 55 & 25 & 0 & 0 \\
 & Theater & 340 & 176 & 0 & 0 \\
 & Twin & 14 & 5 & 8 & 8 \\
 & Cubes & 12 & 4 & 13 & 13 \\
\midrule
\multirow{1}{*}{Social}
 & OverSoul & 14 & 5 & 0 & 0 \\
\midrule
\textbf{Total} & \textbf{31} & \textbf{1937} & \textbf{1276} & \textbf{157} & \textbf{151} \\
\bottomrule
\end{tabular}
\end{table}

    \subsection{Privacy Violation Detection Results}

    We evaluate \sysname's ability to detect privacy violations by comparing violations discovered through automated exploration against those found in manual exploration.
    We record all explorations and network traffic when \sysname explores the 50 ground-truth apps.
    19 of the 50 evaluation apps 
    produce no captured network traffic during either manual or automated exploration.
    Table~\ref{tab:combined-eval} shows the network traffic and violation detection results for the remaining 31 apps.
    Among the 31 apps with captured traffic (8 categories), 15 contain at least one privacy violation from our manual analysis.
    \sysname achieves an overall violation coverage of {96.2\%}, detecting 151 out of 157 ground truth violations across 31 apps with captured traffic, 
    despite only exploring for 20 minutes.

    The lowest-coverage app in our evaluation is \textit{WWidgets} (50.0\%, 2 of 4), where two violations occur in widget-configuration flows that require additional navigation steps not reached within the 20-minute window.
    These results show that \sysname is effective for privacy compliance auditing, as it detects the majority of privacy violations within the first 20 minutes of exploration.
    We expect that with more time, \sysname can cover more violations.
    We present the runtime performance of \sysname in~\autoref{sec:app:runtime}. 
    \ycrv{We also examined the raw captures to quantify interception failures due to MitM settings. We observed no certificate rejections, no undecryptable payloads, and no app exhibiting the behavior characteristic of certificate pinning. \autoref{app:tls} reports the details.}
\looseness=-1

    \subsection{Effectiveness of Auto-exploration}
    \label{sec:eval:baseline}

    A natural question is whether the network requests captured during \sysname's automated exploration genuinely reflect interaction-driven behavior, rather than background traffic the app would emit even when left untouched (e.g., periodic analytics requests).
    To answer this, we compare per-app traffic under two conditions on our 50-app ground-truth set:
    (i) \emph{Auto}, with \sysname performing 20-minute UI exploration (from~\autoref{tab:combined-eval}), and
    (ii) \emph{Idle}, 
    where the same app is launched and left foregrounded but untouched for the same 20-minute duration.
    Both runs have a 20-minute time limit from app launch.
    Note that the Auto will end early if all elements are exhausted.
    To account for early termination we restrict each app's \emph{Auto} run timeframe to its \emph{active interval} (first click $\sim$ last click $+\,3$s) and trim the \emph{Idle} run to a window of equal length.
    All 50 evaluation apps have a (Auto, Idle) pair.

    \ycrv{Manual and automated exploration follow the same protocol and detection logic, systematically triggering every interactive element until none remain, interacting with one representative item per group of same-type elements, and applying the same key-extraction and taxonomy-mapping pipeline. They differ only in how the UI is driven: a human interprets the interface directly, whereas automation relies on non-deterministic OmniParser detection, so the two traverse elements in a different order and depth and do not reach identical states or paths. Manual exploration therefore captures the upper bound of an app's network flows, which is why GT request counts exceed Auto in Table~\ref{tab:combined-eval}.}

\bheading{Results.}
 Of the 50 apps, 26 produce non-zero traffic in both conditions, 5 trigger requests only in \emph{Auto}, 
 and 19 produce no traffic in either condition.
\ycrv{Aggregating across all 50 apps, \sysname's exploration triggers \textbf{1{,}276} unique requests versus \textbf{405} from a length-matched idle baseline, i.e., a $\mathbf{3.15\times}$ increase.
The average network request per minute is 3.51 (Auto) vs. 0.63 (Idle), indicating a $\mathbf{5.58\times}$ boost. }
A detailed per-app comparison is shown in~\autoref{tab:auto-vs-idle} (in~\autoref{sec:app:runtime}).
Among the 26 apps with traffic in both conditions, 17 see strictly more traffic under \emph{Auto} and 7 are tied (e.g., \cc{Kineo}, $1$ vs.\ $1$).
the largest gaps appear in interaction-heavy apps such as \cc{Broadcasts} ($238$ vs.\ $1$, $238\times$) and \cc{alo~Sanctuary} ($102$ vs.\ $1$, $102\times$).

Overall, these results confirm that \sysname's exploration is effective in triggering new traffic that cannot be observed without interaction.
We also present a more detailed study on 1) correlation between clicks and traffic, 2) runtime overhead breakdown of \sysname in~\autoref{sec:app:runtime}.
\ycrv{ We also studied the results of longer runtime (1 hr) in \autoref{app:longrun}.}

\ycrv{A naive metric is the fraction of manually reached UI states that \sysname rediscovers. However, this is not suitable in our case. AVP provides no view hierarchy, so ``state'' is defined by our own perceptual dedup heuristic over detected UI elements (\autoref{sec:sys:comptwo}); there is no ground-truth state identity against which either side can be scored, and the same functional screen can be counted once or several times depending on head pose. Consistent with this, repeated runs of the same app produce noticeably different state counts. We therefore evaluate \sysname on the outcome that is well-defined and that downstream analysis actually consumes (the set of privacy violations recovered) for which manual analysis does provide a ground truth.}

\section{Large-scale Analysis}
\label{sec:measure}

In this section, we present the first large-scale analysis of automated AVP app exploration in the wild to capture privacy violations in network traffic. \sysname explores 324 AVP apps in our large-scale dataset. Of these, 247 apps (76.2\%)   generate network traffic during our exploration window, resulting in a total of 25{,}901 network requests triggered across the experiment; the remaining 77 apps (23.8\%) emit no traffic.

\begin{figure*}[t]
    \centering
    \includegraphics[width=0.8\linewidth]{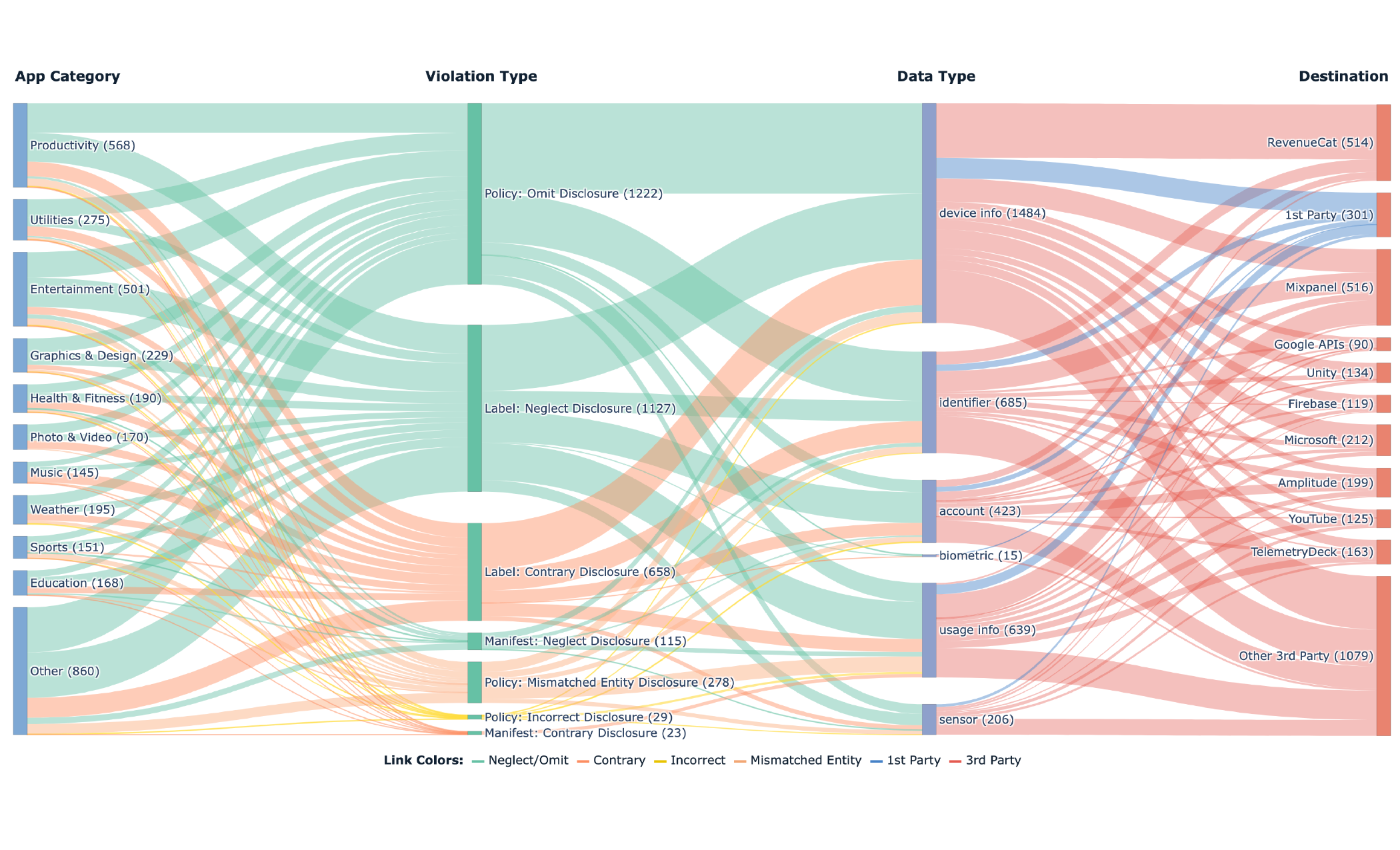}
    \vspace{-30pt}
    \caption{Data flow of all violation traffic.}
    \label{fig:data-flow}
\end{figure*}

\noindent\textbf{\textit{{Violation Landscape.}}}
Our privacy compliance analysis shows that 188 of the 247 apps with captured traffic exhibit at least one violation, and 61.1\% of observed privacy-related data flows occur without proper disclosure. The distribution of all violation traffic is shown in~\autoref{fig:data-flow}; productivity, entertainment, and utility apps are the top three non-compliant app categories.
Across the three disclosure documents, privacy labels account for the largest share of violations: 1,785 label violations (658 contrary, 1,127 neglect) from the 324-app dataset. By comparison, we observe 1,529 policy violations (29 incorrect, 278 mismatched-entity, 1,222 neglect/omit) and 138 manifest violations (23 contrary, 115 neglect).
Furthermore, the most frequently undisclosed data types are device information, identifiers, and usage information, and they are often transmitted to third-party advertising or analytics endpoints such as \textit{Mixpanel}, \textit{RevenueCat}, and \textit{Amplitude} for user tracking, posing significant privacy risks to end users.
Note that our result is a conservative \textbf{lower bound} of potential privacy non-compliance, due to the inherent limitations of dynamic analysis, where the UI exploration may not be exhaustive within a fixed timeframe.

    \subsection{Privacy Policy Non-Compliance}
    \label{sec:policy}
    We collected privacy policies for 247 AVP apps with network traffic and successfully analyzed 212 apps.
    The remaining 35 apps were excluded due to anti-crawl mechanisms, unavailable web content, or non-English content.
    Following prior work~\cite{wang2018guileak,zimmeck2019maps,yu2018ppchecker}, we employed the flow-to-policy consistency model~\cite{andow2020actions} to identify three types of non-compliance (\autoref{fig:violation:policy} in~\autoref{sec:app:heatmap}):

    \vspace{3pt}\noindent$\bullet$ \textbf{\textit{Omit Disclosure}} arises when data is collected, but there are no statements in the privacy policy that disclose or mention this collection. In our study, we identified 1,222 instances of Omitted Disclosure across 137 apps with policy violations.
    The most frequently omitted data types are \textit{device identifiers}, \textit{platform information}, \textit{app version}, and \textit{system version}.
    Notably, the vast majority of omitted disclosures involve data collected by third-party services, while only 9.1\% involve first-party collection, suggesting that developers often fail to account for data collection by embedded SDKs.

    \vspace{3pt}\noindent$\bullet$ \textbf{\textit{Incorrect Disclosure}} is present when data is indeed collected, but the privacy policy conversely states that such data collection does not occur (i.e., a negative sentiment sharing or collection statement). In our study, we identified 29 instances of Incorrect Disclosure across 12 apps with policy violations.
    The most commonly incorrectly disclosed data types are \textit{usage time}, \textit{usage info}, \textit{session data}, and \textit{geographical location}.
    The vast majority of incorrect disclosures originate from third-party services/SDKs, contradicting their ``no data collection'' claims.

    \vspace{3pt}\noindent$\bullet$ \textbf{\textit{Mismatched Entity Disclosure}} arises when data is collected and sent to a specific recipient entity (often a third party, e.g., an advertiser), but the privacy policy attributes that data collection to a different entity (often the first party, i.e., the app developer). 
    This mismatch matters because users may be willing to share data with the app developer but not with unknown or untrusted third parties, so misattributing the recipient can mislead users' privacy expectations and erode trust.
    In our study, we identified 278 instances of Mismatched Entity Disclosure across 81 apps, where the vast majority of violations involved data collected by third-party services without proper attribution. The most common undisclosed third-party collectors include RevenueCat (in-app purchase management), Unity Analytics, Google/Firebase services, and Mixpanel (analytics).
    \looseness=-1

\begin{figure}[t]
    \centering
    \includegraphics[width=.99\linewidth]{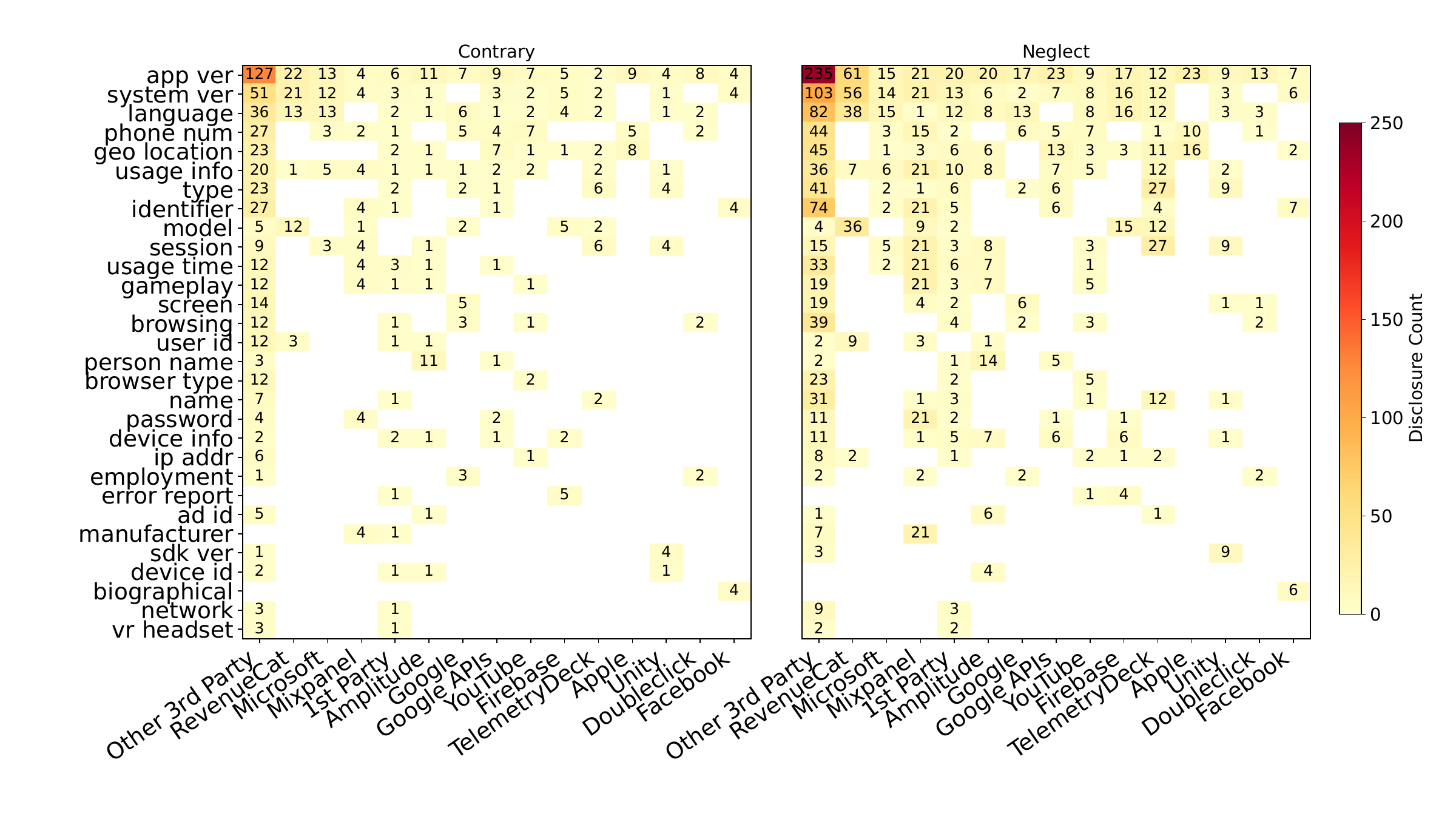}
    \vspace{-10pt}
    \caption{Results: Privacy Label (Partial)}
    \label{fig:violation:label}
\end{figure}

    \subsection{Privacy Label Non-Compliance}
    \label{sec:label}
    We categorize the discovered violations into two primary types based on the nature of the privacy label discrepancy~(\autoref{fig:violation:label}):

\ycrv{The data types in different figures are at different granularity.
\autoref{fig:data-flow} gives an overview of all violation traffic, aggregated
into the six top-level data type categories defined in the VPVet taxonomy.
\autoref{fig:violation:label} details privacy-label violations over the top-30 data
types. \autoref{fig:neglected-datatypes-purpose} and \autoref{fig:purpose-datatype-heatmap} analyze purpose
non-compliance over the 15 top-level Apple privacy-label categories.
Relatedly, \emph{person name} is a fine-grained data type originating from the
VPVet/PoliCheck dataset, whereas \emph{Name} is a coarser-grained type defined
in Apple's privacy-label taxonomy; they refer to the same data category at
different levels of granularity.}
    
    \vspace{3pt}\noindent$\bullet$ \textbf{\textit{Neglect Disclosure}}: Apps that collect data types without any corresponding privacy label declaration. These represent complete omissions where developers fail to disclose data collection practices entirely. 
    In our study, Neglect Disclosure accounts for 63.1\% (1,127 instances) of all privacy label violations, with the most commonly neglected data types being \textit{diagnostic data}, \textit{usage info}, and \textit{device id}.
    See~\autoref{sec:case-studies} for a representative Neglect Disclosure case study.

    \vspace{3pt}\noindent$\bullet$ \textbf{\textit{Contrary Disclosure}}: Apps that explicitly state they do not collect certain data types in their privacy labels while actually collecting that data during runtime. These violations represent direct contradictions between stated and actual privacy practices. In our study, Contrary Disclosure accounts for 36.9\% (658 instances) of violations, affecting 64 apps with a total of 1,423 label-violating network requests.
    Notably, first-party apps exhibit significantly fewer Contrary Disclosure violations compared to third-party SDK-driven collection. This pattern suggests that app developers are generally aware of their own data collection practices, but may not fully understand or account for the data collected by integrated third-party services.
    See~\autoref{sec:case-studies} for a representative Contrary Disclosure case study.
    \looseness=-1

    \subsection{Privacy Manifest Non-Compliance}
    \label{sec:manifest}
    Apple introduced Privacy Manifests (\texttt{PrivacyInfo.xcprivacy}) in 2024, requiring developers to declare privacy-sensitive API usage, tracking domains, and collected data types directly within the app bundle~\cite{Privacy-manifests-specification}. 
    Unlike privacy policies (free-form text) and privacy labels (App Store metadata), privacy manifests are embedded in the IPA file and follow a structured plist format, enabling automated compliance verification.
    
    \bheading{Dataset Coverage.}
    Of the 324 AVP apps in our large-scale dataset, 84 (25.9\%) include a Privacy Manifest.  The declaration patterns cover four key manifest fields. While all 84 apps declare \texttt{NSPrivacyAccessedAPITypes} (required API usage reasons), only 30 (35.7\%) declare \texttt{NSPrivacyCollectedDataTypes}, and a small minority list any \texttt{NSPrivacyTrackingDomains}. This suggests that most developers treat the manifest as an API-compliance checkbox rather than a comprehensive privacy declaration.

We identify manifest-specific violations in 41 of 84 apps (48.8\%), with 138 total violation instances (23 Contrary, 115 Neglect). We categorize these violations using the same two-type framework applied to privacy labels~(\autoref{fig:violation:manifest} in~\autoref{sec:app:heatmap}):

\vspace{3pt}\noindent$\bullet$ \textbf{\textit{Contrary Disclosure (23 apps).}}
Of the 84 apps with manifests, 23 explicitly set \texttt{NSPrivacyTracking} to \texttt{False} yet still transmit privacy data to known tracking domains (e.g., \texttt{mixpanel.com}, \texttt{amplitude.com}, \texttt{doubleclick.net}) during runtime, constituting a direct contradiction.

\vspace{3pt}\noindent$\bullet$ \textbf{\textit{Neglect Disclosure (18 apps).}}
Neglect Disclosure manifests in two forms.
First, among 7 apps that correctly set \texttt{NSPrivacyTracking} to \texttt{True}, 1 provide incomplete \texttt{NSPrivacyTrackingDomains} lists that omit observed tracking endpoints.
Second, 17 apps declare \texttt{NSPrivacyCollectedDataTypes} but omit data types that are actually collected, with the most frequently missing types including identifiers and usage data %
collected by third-party SDKs.
\looseness=-1

\subsection{Non-Compliance due to Third-party SDKs}
\label{sec:sdk}

Motivated by our observation in \autoref{sec:policy}  and \autoref{sec:label} that the majority of privacy violations in our dataset are associated with such third-party data flows,
we focus on data transmitted via third-party SDKs. 
We categorize SDKs into five categories based on manual analysis:
analytics, authentication, advertising, subscription, and others.

Figures~\ref{fig:policy-violations-by-cat-sdk},~\ref{fig:label-violations-by-cat-sdk}, ~\ref{fig:manifest-violations-by-cat-sdk} (\autoref{sec:app:SDK}) shows that 
privacy policy, label and manifest omissions are not evenly distributed across Apple's data taxonomy. The SDK related violations are most prevalent in high-volume categories such as \emph{Productivity} and \emph{Entertainment}, followed by \emph{Utilities}. 
In these categories, \emph{Analytics} SDKs account for the most of the violations.

In contrast, certain categories exhibit a different dominant source.  For example, in \emph{Games}, violations are more heavily driven by
\emph{Advertising} SDKs. While analytics-related SDKs dominate at scale, the relative contribution of SDK types varies across categories, aligning with differences in app usage patterns and monetization strategies.

\ycrv{\autoref{tab:sdk-category} reports how many apps underlie each category,
since violation counts alone conflate category size with per-app severity.
Productivity (26 apps) and Entertainment (30 apps) dominate in absolute terms
because they are the largest categories.}

\begin{table}[t]\centering\footnotesize
\caption{Apps and violations per category}
\label{tab:sdk-category}
\begin{tabular}{lrr}
\toprule
\textbf{Category} & \textbf{Apps} & \textbf{Violations}  \\
\midrule
Productivity        & 26 & 568 \\
Utilities           & 19 & 275 \\
Games               &  2 &  60 \\
Entertainment       & 30 & 501 \\
Lifestyle           &  7 & 140 \\
Health \& Fitness   & 10 & 190 \\
Education           & 10 & 168 \\
Photo \& Video      &  9 & 170 \\
Graphics \& Design  & 12 & 229 \\
Social              &  6 &  56 \\
Weather             &  8 & 195 \\
Sports              &  5 & 151 \\
Music               & 10 & 145 \\
Finance             &  5 & 109 \\
Reference           &  5 &  70 \\
Puzzle              &  4 &  69 \\
Other (13 categories) & 20 & 356 \\
\midrule
\textbf{All} & \textbf{188} & \textbf{3{,}452} \\
\bottomrule
\end{tabular}\end{table}

\begin{figure}[t]
    \centering
    \includegraphics[width=.9\linewidth]{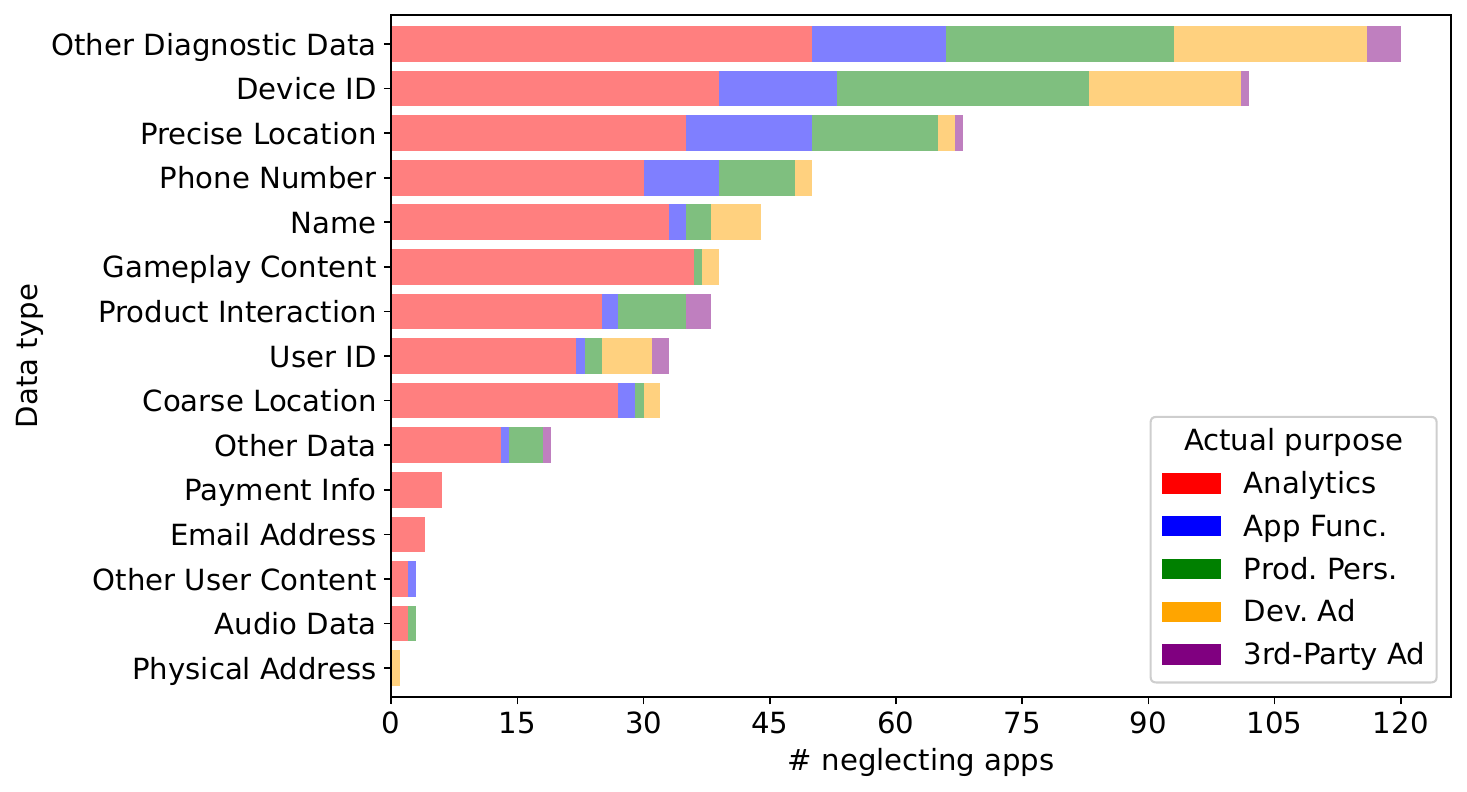}
    \caption{Undisclosed purposes in privacy labels.}
    \label{fig:neglected-datatypes-purpose}
\end{figure}

\begin{figure}[t]
    \centering
    \includegraphics[width=.85\linewidth]{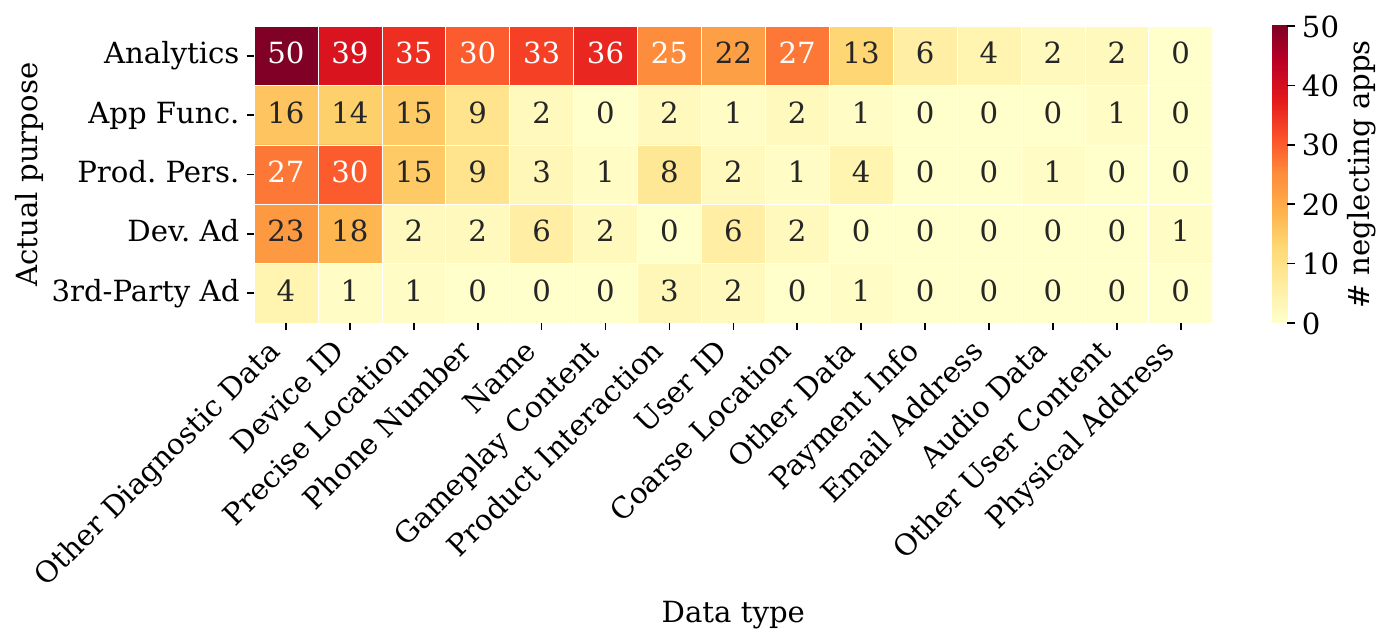}
    \caption{Inadequate disclosures across different purposes.}
    \label{fig:purpose-datatype-heatmap}
\end{figure}

\subsection{Purpose Non-Compliance}
\label{sec:purpose}

Each app's privacy label specifies collected data types and their intended purposes. We assess compliance by comparing the privacy-label purpose for each data type against the inferred runtime purpose of actual data flows; any discrepancy constitutes a \emph{purpose neglect} violation. To infer runtime purposes, we employ the purpose classifier from Lalaine~\cite{xiao2023lalaine}, which predicts Apple's five privacy purposes (\textit{App Functionality}, \textit{Analytics}, \textit{Product Personalization}, \textit{Developer's Advertising or Marketing}, \textit{Third-Party Advertising}) using app metadata and network-level features.

\autoref{fig:neglected-datatypes-purpose} shows that the most common omissions involve \emph{Other Diagnostic Data} and \emph{Device ID}, with further gaps in \emph{Precise Location} and \emph{Phone Number}. Across these prevalent types, \emph{Analytics} emerges as the dominant inferred purpose, suggesting overlooked disclosures often stem from measurement and telemetry infrastructure rather than app-specific logic. Notably, data types such as \emph{Device ID} and \emph{Other Diagnostic Data} are linked not only to Analytics, but also to App Functionality, Product Personalization, and Developer Advertising.
Regardless of data type, collection typically serves multiple purposes, especially \emph{Analytics} and \emph{Product Personalization}, with a substantive \emph{Third-Party Advertising} component. Diagnostic data, while frequently omitted, is widely monetized by analytics and ad-tech SDKs.
\autoref{fig:purpose-datatype-heatmap} further details violations by purpose and data type. The concentration is highest in \emph{Analytics}.
spanning both diagnostic and user-centric data: \emph{Other Diagnostic Data} (50), \emph{Device ID} (39), \emph{Gameplay Content} (36), \emph{Precise Location} (35), \emph{Name} (33), \emph{Phone Number} (30). 
Many data types, such as \emph{Device ID}, support multiple purposes (e.g., Analytics, Product Personalization, Developer Advertising, and App Functionality), revealing that the problem lies not only in whether data types are disclosed, but also in whether their purpose is accurately represented.

\ycrv{Concretely, data that developers file under \emph{Diagnostics} in our traffic includes Crashlytics installation identifiers and OS build versions (\cc{x-crashlytics-installation-id}, \cc{x-crashlytics-os-build-version}), the device model string \cc{RealityDevice14,1} (sent to TelemetryDeck and Crashlytics), SDK and app versions (\cc{sdk\_ver=2022.3.47f1} to Unity; \cc{\$app\_version\_string} to Mixpanel), screen resolution, and network type (\cc{networkType=wifi}). Data filed under \emph{Usage Data} includes session identifiers (\cc{\$mp\_session\_id} to Mixpanel; \cc{sessionID} to TelemetryDeck) and per-event timestamps. These examples illustrate that ``diagnostics'' and ``usage'' collection routinely carries stable identifiers suitable for cross-app linkage. More broadly, \sysname operationalizes several goals of the NIST Privacy
Framework~\cite{nist-privacy-framework} for AVP apps: identifying actual data
processing through runtime traffic, comparing observed flows with disclosed
privacy practices, and helping developers govern third-party SDK behavior. It
thus serves as a practical privacy-auditing tool that bridges abstract
privacy-risk goals with concrete evidence of what data leaves AVP apps and
whether those flows are properly disclosed.}

\subsection{Responses from Apple and Developers}
\label{sec:discussion:responses}

Following responsible disclosure practices, we reported our findings to Apple and app developers. 
We summarize their responses as follows.

\bheading{Apple's Responses.}
We reported our findings to Apple via the Feedback Assistant and App Store Review channels, providing affected apps, supporting evidence, and a summary of the most prevalent violations.
\ycrv{After receiving our reports, Apple acknowledged, validated, and reproduced our findings.}

\bheading{Developer Responses.}
\ycrv{We searched for developer contact information from App Store information and developer websites, and were able to find the contact information for 119 apps out of 188 apps with violations. We reported our findings to them, and we have received 10 responses so far.}
We categorize the responses into four categories (some of them belong to multiple categories):

\emph{(i) Acknowledging disclosures.}
Two developers agreed to update or refine their privacy disclosures. For example, \emph{Alpenglow Sunset Predictions} acknowledged that User ID transmission (via RevenueCat and Unleash SDKs) was previously underdisclosed, and revised the privacy label to declare User ID under ``Analytics / Not Linked to You''. 
Similarly, \emph{Space Vision} stated that its existing ``Crash Data -- App Functionality'' label covers Firebase Crashlytics, but committed to making the disclosure more precise in the next release.

\emph{(ii) Relying on SDK documentation or SDK-provided semantics.}
Four developers based their disclosures on SDK documentation or information provided by SDK vendors regarding data semantics. For instance, \emph{1Blocker} cited RevenueCat's official privacy documentation to support its current label assignments, and 
\emph{Alpenglow} argued that RevenueCat's use of Apple Search Ads attribution leverages Apple's AdServices framework rather than persistent device identifiers. 
These responses suggest that developers often defer to SDK vendors' data-handling descriptions, rather than directly verifying runtime data flows.

\emph{(iii) Arguing that existing labels already cover observed behavior.}
Five developers stated that their current privacy-label disclosures adequately cover the reported data flows. 
For example, \emph{Space Vision} mapped Firebase Crashlytics to ``Crash Data -- App Functionality,'' and \emph{Spatial Physics Playground} contended that its first-party requests were already labeled as ``Diagnostic Data, Not Linked to Identity.'' 
Similarly,
\emph{Sequel Media Tracker} stated that its privacy label already includes purchases, location, identifiers, and usage data. 
These cases highlight the broad and sometimes ambiguous categorization permitted by Apple's privacy-label taxonomy.

\emph{(iv) Disputing data-type classification or reported flows.}
Four developers contested our traffic analysis or the classification of specific data types. For example, \emph{Alpenglow} argued that an attribution token should not be classified as a persistent Device ID; \emph{Cardhop} requested clarification of what ``Browsing'' represented.
This indicates that the classification of data types is not always clear-cut, and there are cases where developers may have different interpretations of the data types.

\ycrv{In the disputed cases, developers often categorized specific data items
into broader labels such as ``Diagnostics'' or ``Usage Data,'' even when the
transmitted fields more precisely matched Apple-defined categories such as
identifiers or browsing activity. For example, one developer argued that a
Device ID sent to Mixpanel was covered under ``Usage Data,'' but Apple requires
Device ID to be declared as such even when marked as not linked to identity.
Our categorization follows Apple's published data-type definitions and reflects
how these items should be disclosed to end users.}

\subsection{Summary of Findings}

We summarize our findings from large-scale analysis as follows.

\bheading{Prevalence of Non-Compliance.}
188/324 (58.0\%) AVP apps violate one or more privacy disclosure requirements and 61.1\% of observed privacy-related data flows occur without proper disclosure. This is a conservative lower bound, as we only performed 20-minute exploration.

\bheading{Policy, Label and Manifest Non-Compliance.}
Omissions and inaccuracies are common in both privacy policies and App Store privacy labels.
Adoption of privacy manifests remains inconsistent and frequently superficial; nearly half of apps with manifests omit required elements or fail to accurately represent observed data flows.

\bheading{SDKs as Primary Compliance Bottleneck.}
Third-party SDKs account for the majority of undisclosed network traffic, typically outside the direct control or awareness of app developers.

\bheading{Purpose Misalignment.}
Substantial discrepancies persist between declared purposes and actual runtime data usage, undermining the integrity of App Store privacy disclosures.

\bheading{Developer Uncertainty.}
Ambiguities in Apple's data taxonomy, coupled with insufficient tooling for SDK attribution, contribute to confusion and inconsistent compliance among developers.

\ycrv{\textbf{AVP-specific leakage.}
AVP-specific data account for only a small fraction of observed violations. Of 3{,}452 violations, 46 (1.3\%) map to AVP-specific taxonomy nodes. Manual inspection confirms that none of the 46 contains raw XR sensor measurements. Importantly, our static analysis of all 374 app binaries shows that 54 declare an XR sensor usage string and 39 link an authorization-gated ARKit provider, indicating that such capabilities are present in the ecosystem, yet we observe no corresponding raw sensor transmission. This finding is consistent with visionOS's privacy architecture: Apple states that the system mediates camera, sensor, gaze, and hand inputs without directly exposing them to apps, while more sensitive providers require explicit authorization in restricted immersive contexts~\cite{apple-visionos-privacy}. Thus, the absence of raw XR telemetry provides encouraging empirical evidence that visionOS's system-mediated sensing boundary limits direct network exposure of raw sensor inputs.}

\section{Discussion}
\label{sec:discuss}
We discuss the implications and recommendations. We further discuss our limitations and future work in~\autoref{sec:app:more-discuss}. 

\ycrv{\bheading{Implications.} AVP-specific sensing data can reveal physical
behavior and surrounding context. Even where such streams remain on-device, the
identifiers that are sent through network (e.g., User ID and Device ID shared with third
parties) enable cross-app linkage and user profiling, and the small,
distinctive AVP install base makes that linkage easier than the equivalent on a
phone.}

\ycrv{Quantitatively, our non-compliance rate is comparable to  
OVRSeen's Meta Quest findings (58.0\% of AVP apps with at least one violation vs.\ 58.6\% on Quest). VisionOS providing richer privacy disclosures and manifests, but these additional disclosure mechanisms do not translate into more consistent compliance. We attribute this to two factors. First, the AVP ecosystem is still relatively immature, and developers may not yet have established reliable practices for accurately documenting the privacy data. Second, visionOS is a closed platform that provides limited visibility into application behavior: without root access or third-party auditing tools, developers have few practical ways to independently verify how privacy data flows through their applications. AVP-INSPECT addresses this gap by providing an external method to audit these flows.}

\bheading{Recommendations.}
Based on our findings, we offer recommendations for three stakeholder groups in the AVP ecosystem:

\noindent\emph{For AVP Users.}
    Users should not rely solely on App Store privacy labels or manifest disclosures, as our findings show these are frequently incomplete or inaccurate. Users are advised to routinely review and manage app permissions, restrict access to sensitive device features, and uninstall or avoid apps with a history of privacy violations. Where feasible, users should favor applications from developers with strong, transparent privacy practices.

\noindent\emph{For the AVP Platform (Apple).}
    The current self-reporting framework for privacy labels and manifests is inadequate for ensuring compliance. Apple should implement automated, ongoing runtime verification of privacy declarations, and systematically audit both in-house and third-party SDKs for undisclosed data collection. Policy reforms should require explicit data usage declarations for all bundled SDKs, enforce penalties for repeated non-compliance, and provide users with increased transparency, such as in-app notifications about ongoing data flows and results from independent privacy audits. For example, Apple can provide an option for users to see the ongoing data flows and the related privacy-sensitive data categories of the foreground app.
\ycrv{Concretely, we recommend three classes of actions.
\emph{1) Documentation}: provide a finer-grained AVP-specific taxonomy for
spatial and biometric disclosures, and treat \cc{x-platform: visionOS} sent to
third parties as tracking-relevant in privacy manifests.
\emph{2) Enforcement}: surface SDK-level privacy manifests to developers, and
cross-check runtime traffic against machine-readable manifests during app
vetting. \emph{3) APIs}: add distinct runtime permissions and persistent
indicators for spatial and biometric sensing APIs.}

\noindent\emph{For AVP Developers.}
    Developers must perform comprehensive audits of all integrated SDKs and explicitly disclose all collected data types and purposes in both App Store privacy labels and manifests. Automated privacy compliance checks should be integrated into the development lifecycle, with disclosures updated promptly in response to any code or SDK changes. Developers are responsible for ensuring that privacy documentation remains accurate and up-to-date, reflecting the actual behavior of both first- and third-party code within their apps.
    \ycrv{Concretely, we recommend four classes of actions: 1) audit third-party SDKs through
runtime traffic inspection; 2) disclose third-party recipients and purposes, not
only first-party collection; 3) keep privacy policies, labels, and manifests
synchronized; and 4) update all privacy documents whenever the app or an embedded SDK
changes.}

\section{Related Work}
\label{sec:related}

We discuss prior research on XR privacy analysis and XR dynamic testing. Additional related work on XR privacy attacks and defenses, as well as iOS privacy analysis, is covered in \autoref{sec:app:related}.

\bheading{XR privacy analysis.}
Ovrseen~\cite{trimananda2022ovrseen}
is the first work to study inconsistencies between network traffic and privacy policies of VR applications on Oculus (Meta) platforms.
VPVet~\cite{zhan2024vpvet}
is a recent tool for vetting privacy policies across 10 mainstream VR platforms.
Guo~\etal~\cite{guo_empirical_2024} have analyzed 500 VR apps to study their security and privacy issues, and 
Simhadri~\etal~\cite{vamsi_ccs2025} performed a longitudinal analysis on 300+ firmware from Quest and Pico devices.
Brunskong~\etal~\cite{brunskog2025network} presented a network traffic analysis associated with child and adult accounts of six Meta Quest applications,
raising concerns about insufficient  age-appropriate data practices  on VR platforms.
Different from those works which focus on Meta devices,
we perform the first privacy analysis on AVP devices based on network traffic using automated testing.

\bheading{Apple privacy labels.}
\ycrv{Prior work has studied Apple privacy labels from user, developer, and ecosystem perspectives, showing that users may struggle to interpret labels~\cite{zhang2022usable,tahaei2023stuck}, developers face challenges mapping app and SDK behavior to Apple’s data categories~\cite{li2022understanding, alsahdi2026because,xiao2024measuring}, and labels are not always updated as app behavior changes~\cite{li2022understanding}. 
Prior work has also studied inconsistencies between privacy labels and privacy policies~\cite{jain2023atlas}, as well as inconsistencies between iOS mobile app traffic and privacy labels using either passive/no-touch traffic collection~\cite{koch2022keeping} or UI-driven dynamic analysis~\cite{xiao2023lalaine}.
Our work complements this line of research by extending privacy-label compliance analysis from traditional mobile apps to Apple Vision Pro apps. 
Our results further show that disclosure inconsistencies are also prevalent in the AVP ecosystem. We attribute this in part to two factors: the relative immaturity of the AVP ecosystem, and the closed nature of the platform. Without root access or mature third-party auditing tools, developers cannot easily verify what their own apps and embedded SDKs transmit, which is the gap \sysname fills.}

\bheading{XR dynamic testing.}
\ycrv{
Prior XR testing work targets platforms with source access or standard testing suites. White-box and model-based tools such as VRTest~\cite{wang2022vrtest}, VRGuide~\cite{wang2023vrguide}, and VRExplorer~\cite{zhuvrexplorer} automate scene exploration on Unity/SteamVR; Youkai~\cite{figueira2022youkai} and VR-ReST~\cite{correa2018automated} address unit testing and requirements-driven test generation. Other efforts detect visual defects such as stereoscopic inconsistency~\cite{li2024less} and AR object-placement issues~\cite{rafi2022predart}, and Gu~\etal~\cite{gu2025software} survey this landscape. Recently, AutoVR~\cite{kimautovr} was introduced to test Unity-based VR 
applications and detect sensitive data exposure using Frida. Different from our work,
these approaches 1) target platforms (e.g., SteamVR) with standard 
testing suites, 2) rely on source code, root capability, or use dynamic 
instrumentation tools, which are not applicable to AVP due to the 
challenges mentioned in \autoref{sec:overview}.}

\ycrv{\bheading{Non-intrusive GUI testing.} 
RoboTest~\cite{yu2024robotest} is closest to our \textsc{Controller}: it physically actuates smartphones with a robotic arm and recovers widgets from screenshots. We share its non-intrusive, vision-based premise, but AVP has no touch surface and no absolute screen-to-actuator mapping. \sysname{} therefore uses a low-cost hybrid controller: Bluetooth HID for pointer/keyboard input and servo motors for hardware buttons. Its control problem is also different: AVP cursor motion is relative and depth-dependent, requiring feedback-guided binary search and perceptual state deduplication under head-pose variation. Finally, RoboTest targets crash/compatibility bugs, whereas \sysname{} audits privacy compliance from network traffic.}

\section{Conclusion}
\label{sec:conclude}

In this paper, we presented \sysname, a novel testing framework that enables automated privacy analysis of Apple Vision Pro applications, without requiring root access or application source code. We first evaluated \sysname using a 50-app ground-truth dataset to demonstrate its effectiveness, and we performed a large-scale analysis on 324 AVP applications, which reveal that 188 apps have non-compliance issues. 
We have reported our findings to Apple and app developers, and we offer recommendations to help them mitigate such issues.
Our findings highlight the need for improved privacy enforcement mechanisms in emerging XR platforms and provide a foundation for future privacy research in immersive computing environments.

\section*{Acknowledgment}
The authors from George Mason University (GMU) are supported
in part by 1) a seed funding and GRA awards from the CAHMP
(now CHAIS) Center at GMU, and 2) a seed funding from 4-VA,
a collaborative partnership for advancing the Commonwealth of
Virginia.
Yue Xiao is supported in part by the Commonwealth Cyber Initiative (CCI-HC-2Q26-037).

{
\bibliographystyle{ieeetr}
\bibliography{paper}
}

\appendix
\newpage
\section{Ethical Considerations}
\label{sec:ethics}

\bheading{Stakeholders.}
Our study involves three primary stakeholder groups.
First, AVP users rely on privacy disclosures to make informed decisions about app installations and data sharing.
Second, app developers may face scrutiny from identified privacy violations, though our analysis focuses on systemic compliance issues rather than individual fault attribution.
Third, Apple as the platform operator may face increased attention regarding privacy enforcement mechanisms, which we view as an opportunity to strengthen user protection.

\bheading{Data Collection.}
We collected publicly available information from the App Store, including app metadata, privacy manifests, privacy labels, and privacy policies.
Our data collection process was designed to minimize disruption to the platform by implementing rate limiting.
We did not access any user data or private information during our study.

\bheading{Network Traffic Analysis.}
We analyzed network traffic generated during automated app exploration.
All traffic was captured from apps we installed on our own AVP device.
We did not intercept or analyze traffic from other users.
The analysis focused on identifying privacy-related data transmissions to assess compliance with disclosed privacy practices.

\bheading{Researcher Safety.}
Authors who tested AVP applications during the study did so voluntarily.
Testing sessions were limited in duration and discontinued immediately upon experiencing discomfort.

\bheading{Responsible Disclosure.}
We are in the process of reporting our findings to Apple and affected app developers to allow them to address identified privacy violations before public disclosure.
\ycrv{We disclosed our findings before submission. We reported aggregate results
to Apple through the Feedback Assistant and App Store Review channels, and
contacted 119 developers. No user data
was collected; all traffic originates from apps installed on our own device
under our own accounts.}

\section{Open Science}
\label{sec:open-source}

Our artifact can be found at: \url{https://github.com/SECSAT-LAB-GMU/AVP-Inspect} 

\section{Generative AI Usage}
AI tools (e.g., ChatGPT) have been used to check grammar and polish the sentences.

\begin{algorithm}[t]
\caption{App State Exploration Algorithm}
\label{alg:app-exploration}
\scriptsize
\SetAlgoLined
\LinesNumbered

\SetKwProg{Fn}{Function}{:}{}
\SetKwFunction{Main}{Main}
\SetKwFunction{CheckIfStateIsKnown}{CheckIfStateIsKnown}
\SetKwFunction{CheckCurrentState}{CheckCurrentState}
\SetKwFunction{ResumeExploreState}{ResumeExploreState}
\SetKwFunction{JumpToCurrentState}{JumpToCurrentState}
\SetKwFunction{Screenshot}{Screenshot}
\SetKwFunction{Click}{Click}
\SetKwFunction{TryClickBackButton}{TryClickBackButton}
\SetKwFunction{RestartApp}{RestartApp}
\SetKwFunction{GetElements}{GetElements}

\Fn{\CheckIfStateIsKnown{state}}{
    \If{state $\in$ G.nodes}{
        \KwRet true\;
    }
    \KwRet false\;
}

\Fn{\CheckCurrentState{}}{
    new\_state $\leftarrow$ \GetElements{\Screenshot{}}\;
    \eIf{\CheckIfStateIsKnown{new\_state}}{
        \ResumeExploreState{new\_state}\;
    }{
        G.nodes.append(new\_state)\;
        G.edges.append(Button(id, pre\_state, new\_state))\;
    }
}

\Fn{\ResumeExploreState{state}}{
    \While{state.unexplored\_buttons $\neq \emptyset$}{
        button $\leftarrow$ state.unexplored\_buttons.pop()\;
        \Click{button}\;
        \CheckCurrentState{}\;
    }
    \TryClickBackButton{}\;
    \CheckCurrentState{}\;
}

\Fn{\JumpToCurrentState{target\_state}}{
    \RestartApp{appid}\;
    \ForEach{button $\in$ GetPath(first\_state, target\_state)}{
        \Click{button}\;
    }
}

\BlankLine
\Fn{\Main{}}{
    first\_state $\leftarrow$ initial state of the app\;
    G.nodes.append(first\_state)\;

    \ResumeExploreState{first\_state}\;

    \While{$\exists$ state $\in$ G \text{ with } state.unexplored\_buttons $\neq \emptyset$}{
        state $\leftarrow$ next state with unexplored buttons\;
        \JumpToCurrentState{state}\;
        \ResumeExploreState{state}\;
    }
}

\end{algorithm}
\begin{algorithm}[t]
\caption{Iterative Discovery of New Data Types Using Embeddings}
\label{alg:embedding-discovery}
\scriptsize
\KwIn{Unmapped data phrases $D_{unmapped}$, similarity threshold $\theta = 0.8$}
\KwOut{Set of new data types $N$}

\BlankLine

$E \gets$ GenerateEmbeddings($D_{unmapped}$)\;
$N \gets \emptyset$ \tcp{New data types}
$D_{remaining} \gets D_{unmapped}$\;

\BlankLine

\While{$D_{remaining} \neq \emptyset$}{
    \tcp{Manually select a representative data type}
    $t \gets$ ManuallySelect($D_{remaining}$)\;
    $N \gets N \cup \{t\}$\;
    
    \BlankLine
    
    \tcp{Calculate similarity and filter similar phrases}
    $D_{similar} \gets \emptyset$\;
    \ForEach{phrase $d \in D_{remaining}$}{
        \If{Similarity($d$, $t$) $\geq \theta$}{
            $D_{similar} \gets D_{similar} \cup \{d\}$\;
        }
    }
    
    \BlankLine
    
    \tcp{Remove filtered phrases from remaining set}
    $D_{remaining} \gets D_{remaining} \setminus D_{similar}$\;
    
    \BlankLine
    
    \If{no new data type can be manually identified}{
        \textbf{break}\;
    }
}

\BlankLine

\Return{$N$}\;

\end{algorithm}

\section{Algorithms}
\label{app:algorithm}

We present the two algorithms:
1) App state exploration (\autoref{sec:sys:comptwo});
2) Discovery of new data types (\autoref{sec:sys:compthree}).

\section{Taxonomy of \sysname}
\label{app:taxonomy}

\begin{figure}[p]
    \centering\includegraphics[height=1.2\textwidth]{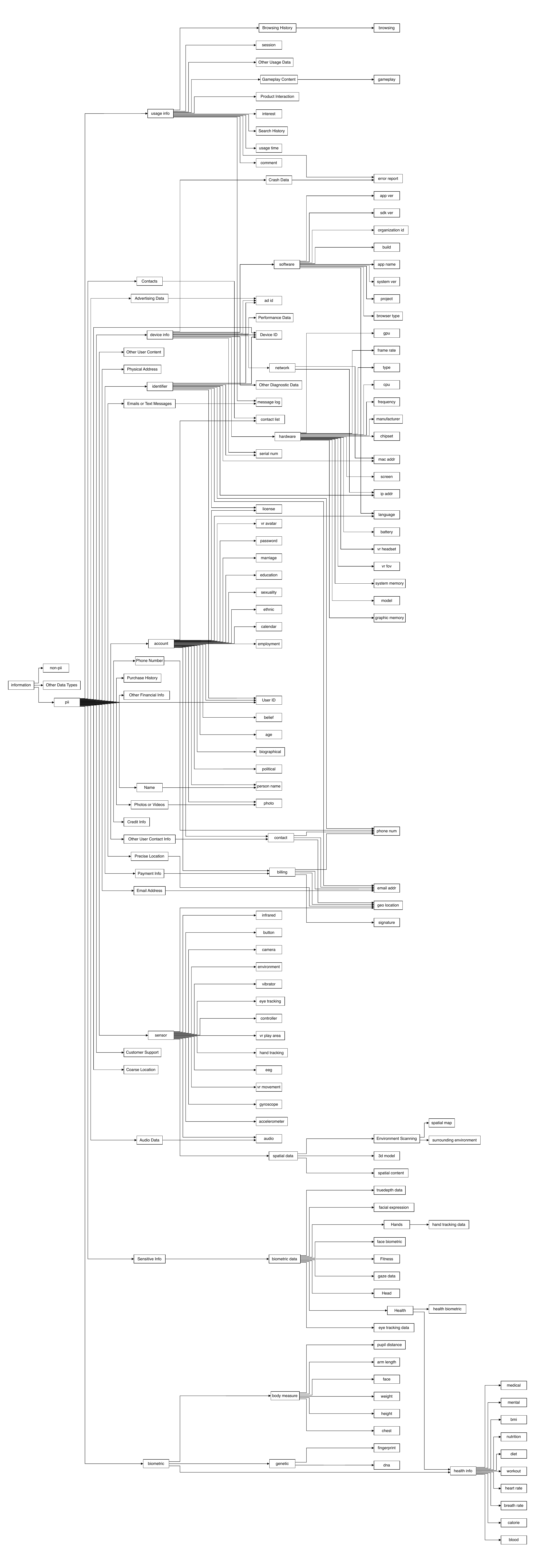}
    \centering
  
    \caption{Taxonomy}
    \label{fig:taxonomy}
\end{figure}

\autoref{fig:taxonomy} shows the taxonomy in full. 

\section{Runtime Performance of \sysname}
\label{sec:app:runtime}

\begin{table}[t]
\centering
\caption{Per-app per-min unique network request comparison: \sysname's exploration (\textbf{Auto}) vs.\ idle baseline (\textbf{Idle}) for all 31 evaluation apps with captured traffic.}
\label{tab:auto-vs-idle}
\footnotesize
\setlength{\tabcolsep}{3pt}
\begin{tabular}{l l r r}
\toprule
\textbf{Category} & \textbf{App} & \textbf{Auto/min} & \textbf{Idle/min} \\
\midrule
\multirow{6}{*}{Productivity}
 & Rad Timer & 0.13 & 0.04 \\
 & FreelanceKit & 0.21 & 0.14 \\
 & Cardhop & 0.12 & 0.05 \\
 & Focus - Timer & 0.00 & 0.00 \\
 & FocusBeats & 0.39 & 0.18 \\
 & Flippy & 0.14 & 0.14 \\
\midrule
\multirow{4}{*}{Utilities}
 & Craft & 1.80 & 0.36 \\
 & Broadcasts & 10.82 & 0.05 \\
 & Qlone & 0.10 & 0.05 \\
 & WWidgets & 0.28 & 0.05 \\
\midrule
\multirow{2}{*}{Games}
 & Retrogram & 0.81 & 0.31 \\
 & Things & 2.50 & 7.85 \\
\midrule
\multirow{4}{*}{Entertainment}
 & Callsheet & 119.60 & 3.12 \\
 & Turn Off the Lights & 0.00 & 0.00 \\
 & Wet Your Beak & 1.09 & 0.63 \\
 & Paramount+ & 12.58 & 0.31 \\
\midrule
\multirow{5}{*}{Lifestyle \& Health}
 & alo Sanctuary & 4.64 & 0.05 \\
 & Decathlon USA & 0.41 & 0.05 \\
 & Mindr & 0.06 & 0.04 \\
 & Bible & 1.89 & 0.18 \\
 & Best Buy & 2.51 & 0.14 \\
\midrule
\multirow{4}{*}{Education}
 & Chemistry & 0.69 & 0.23 \\
 & Drawing Desk & 0.71 & 0.00 \\
 & Foxar & 3.75 & 0.36 \\
 & Inviewer & 0.80 & 0.14 \\
\midrule
\multirow{5}{*}{Creative}
 & Kineo & 0.14 & 0.05 \\
 & Spatial Station & 2.36 & 0.00 \\
 & Theater & 26.25 & 3.46 \\
 & Twin & 0.57 & 0.14 \\
 & Cubes & 0.40 & 0.14 \\
\midrule
\multirow{1}{*}{Social}
 & OverSoul & 0.70 & 0.00 \\
\midrule
\textbf{Total} & \textbf{31} & \textbf{3.51} & \textbf{0.63} \\
\bottomrule
\end{tabular}
\end{table}

\begin{figure*}[t]
    \centering
    \includegraphics[width=.9\textwidth]{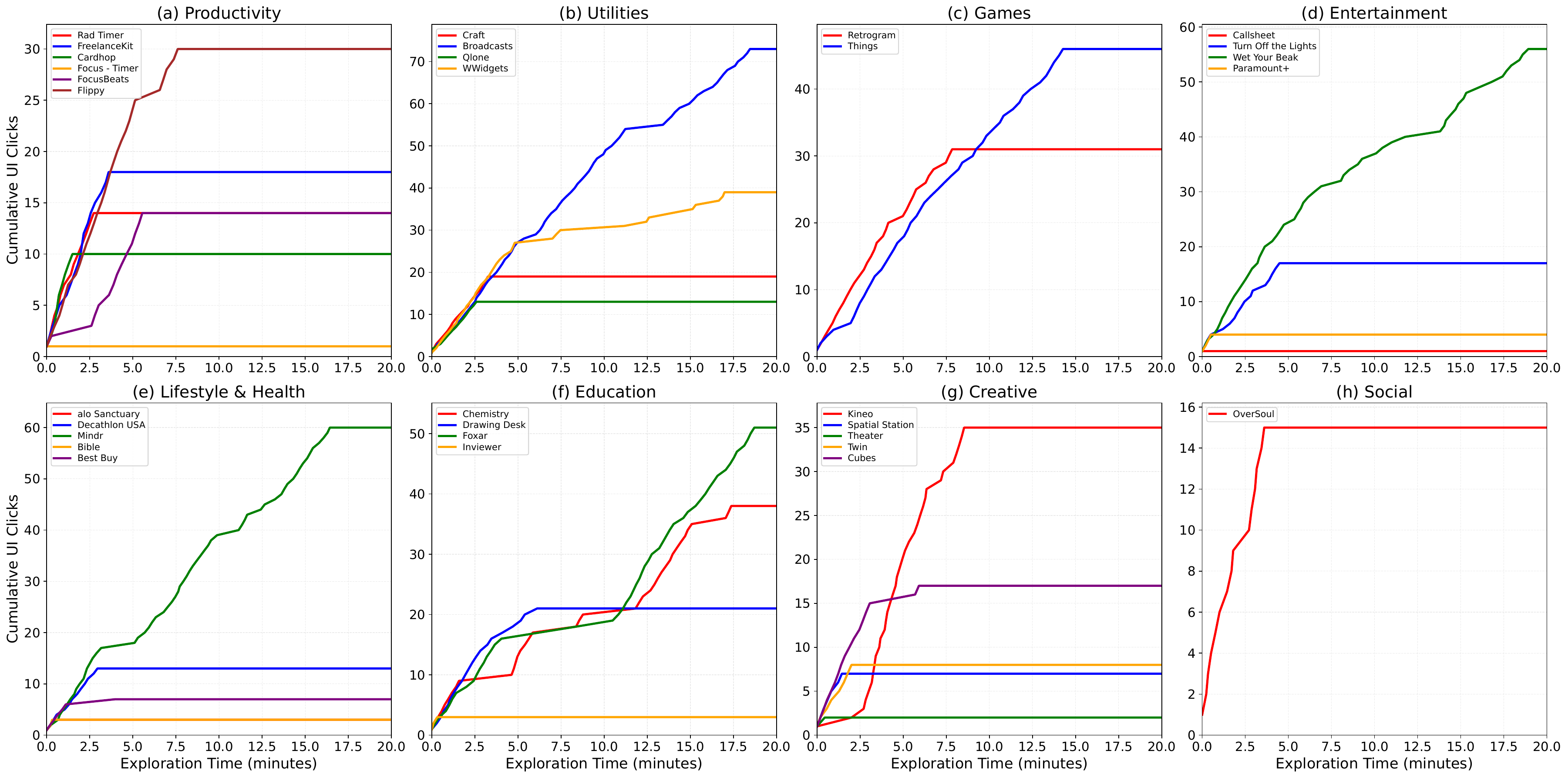}
    \caption{Cumulative UI clicks over time during app exploration, grouped by app category.}
    \label{fig:timeseries_ui_clicks}
\end{figure*}

\begin{figure*}[t]
    \centering
    \includegraphics[width=.9\textwidth]{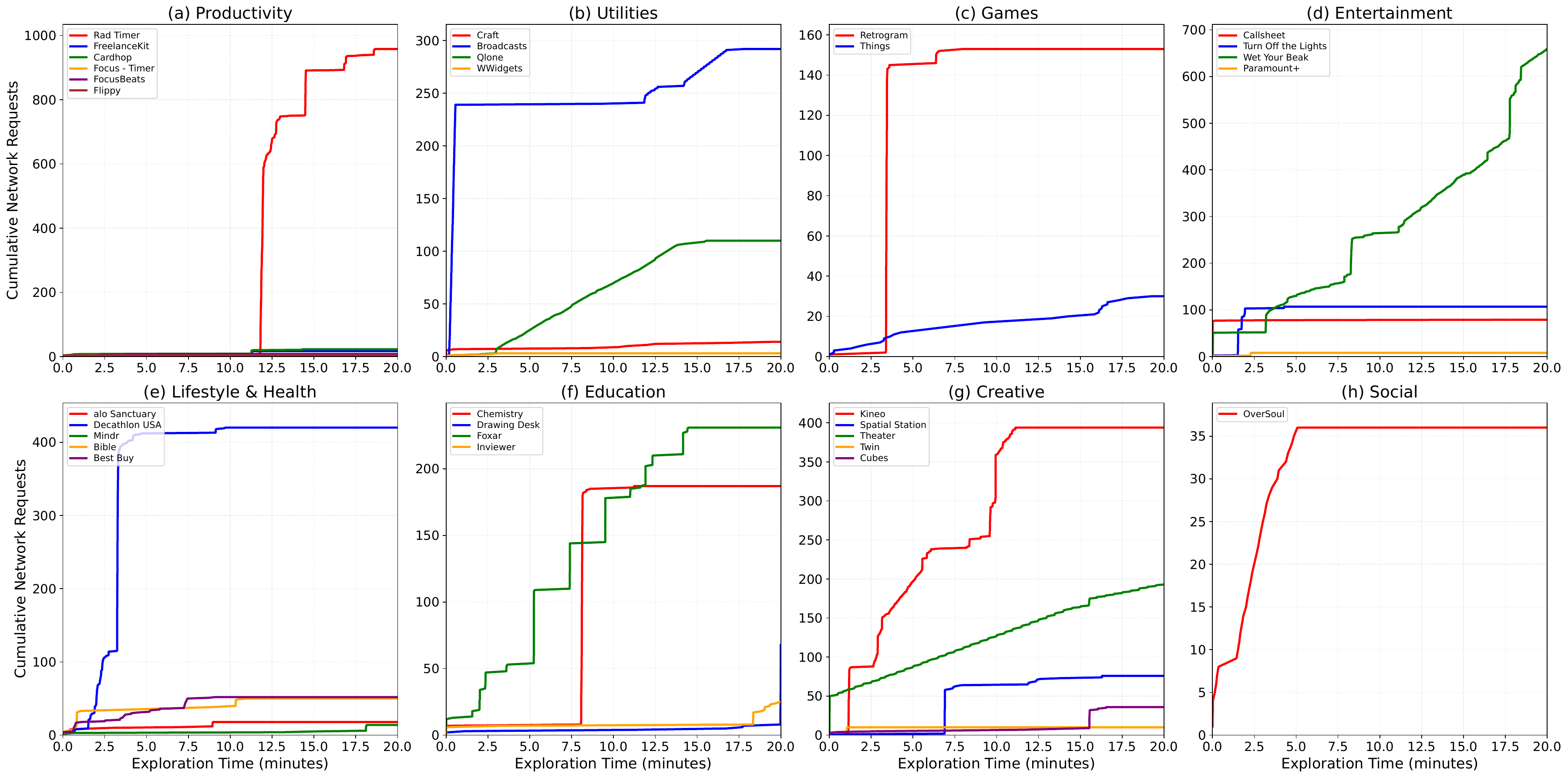}
    \caption{Cumulative network requests over time during app exploration, grouped by app category.}
    \label{fig:timeseries_network_traffic}
\end{figure*}

\bheading{Exploration efficiency.}
\autoref{fig:timeseries_ui_clicks}
and \autoref{fig:timeseries_network_traffic}
show the cumulative UI clicks and
network requests over 20-minute exploration sessions, grouped by app category.
For comparability, both figures use the same subset of ground-truth apps that generated observable network traffic during exploration.
Exploration patterns vary significantly across categories.
Entertainment  apps show consistent patterns, with apps like \emph{Wet Your Beak}, achieving over 40 clicks due to well-structured UIs, and network traffic closely follows click pattern.
In Utilities, \emph{Broadcasts} shows the deepest UI exploration: its click count keeps increasing throughout the 20-minute session and reaches more than 70 clicks.
In Education, \emph{Foxar} generate substantial network traffic generate substantial network traffic relative to clicks, and it triggers multiple asset fetching per interaction.

\begin{figure}[t]
    \centering
    \includegraphics[width=.95\columnwidth]{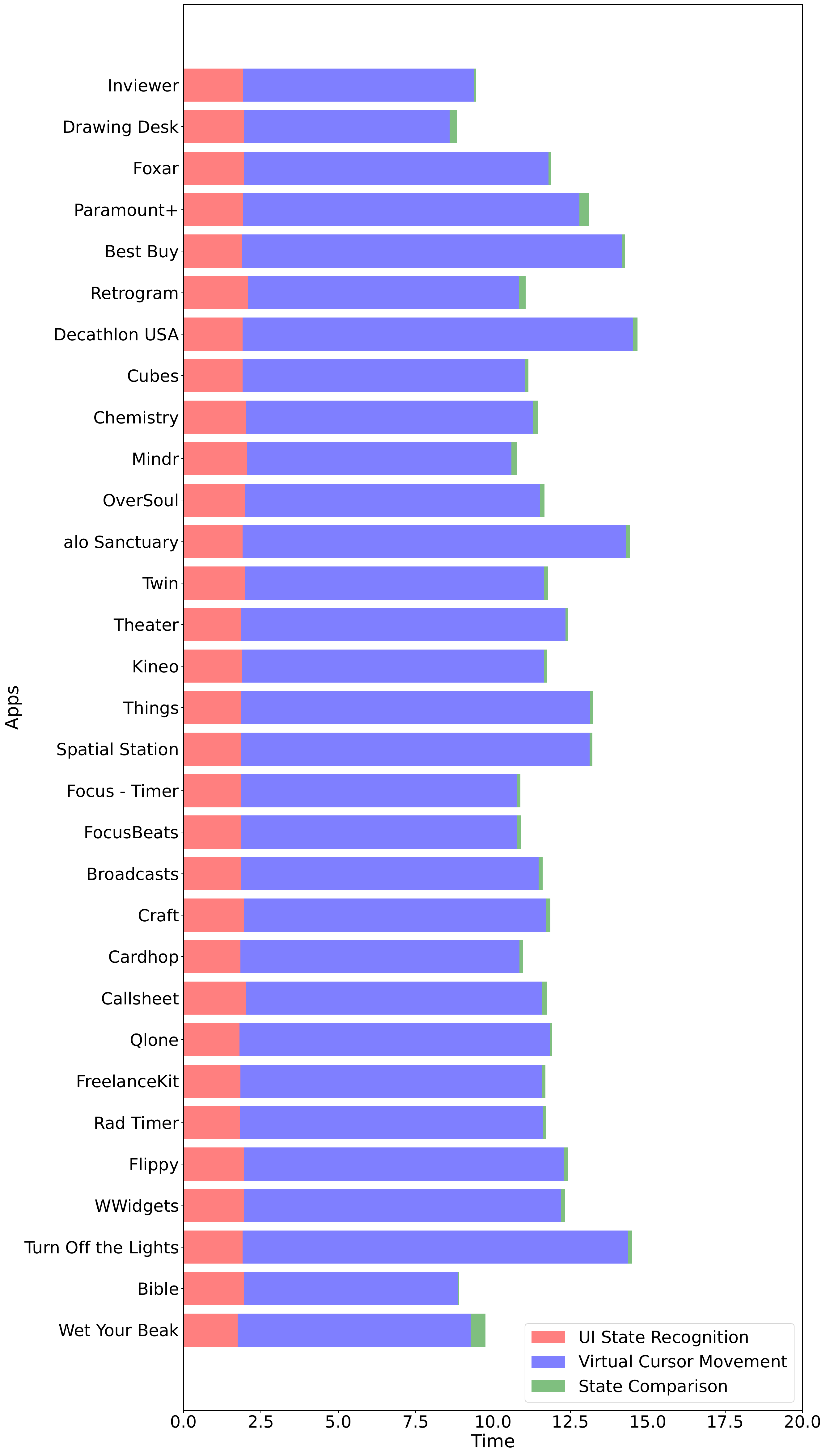}
    \caption{Average time per click for 31 ground-truth apps.}
    \label{fig:time_breakdown}
\end{figure}

\bheading{Time breakdown.}
To understand the time cost of each exploration cycle, we analyze the breakdown of average time spent per click.
As shown in~\autoref{fig:time_breakdown}, each click cycle consists of three parts: 
1) UI state recognition by OmniParserV2,
2) virtual cursor movement,
3) state comparison to check whether a new state is reached.
On average, Each click takes approximately 12.10s. 
The  virtual cursor movement (9.76s) dominates due to the iterative visual calibration required for precise targeting, which is a necessary overhead for our hardware-based approach that works with any app without system modifications or root access.

\begin{figure*}[t]
    \centering
    \includegraphics[width=.99\linewidth]{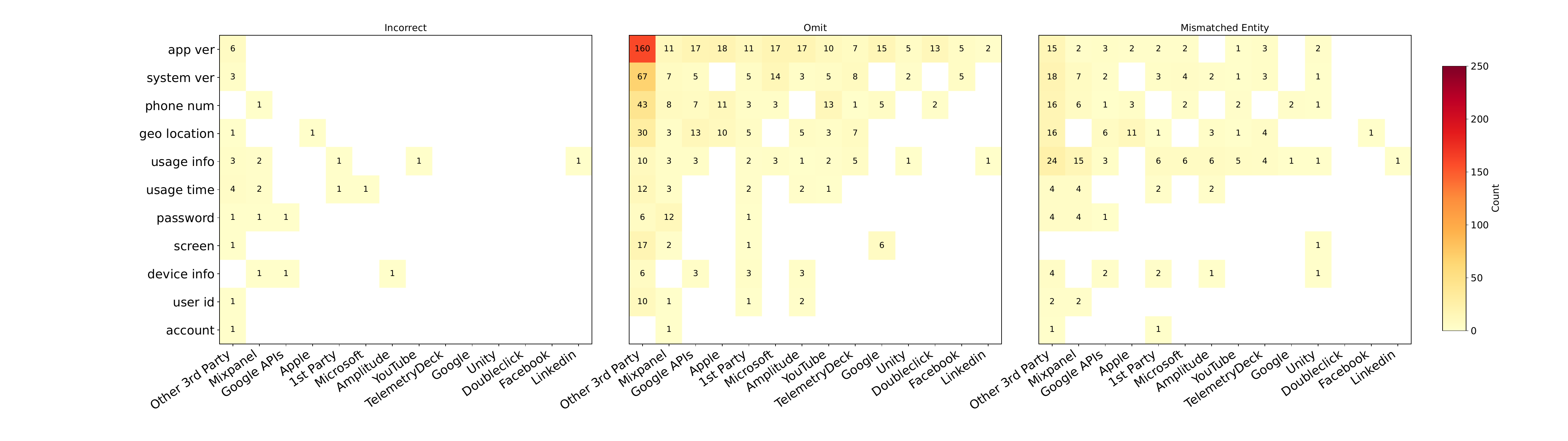}
    \caption{Results: Privacy Policy (Partial)}
    \label{fig:violation:policy}
\end{figure*}

\section{Large-Scale Analysis: Heatmaps}
\label{sec:app:heatmap}

\begin{figure}[t]
    \centering
    \includegraphics[width=.99\linewidth]{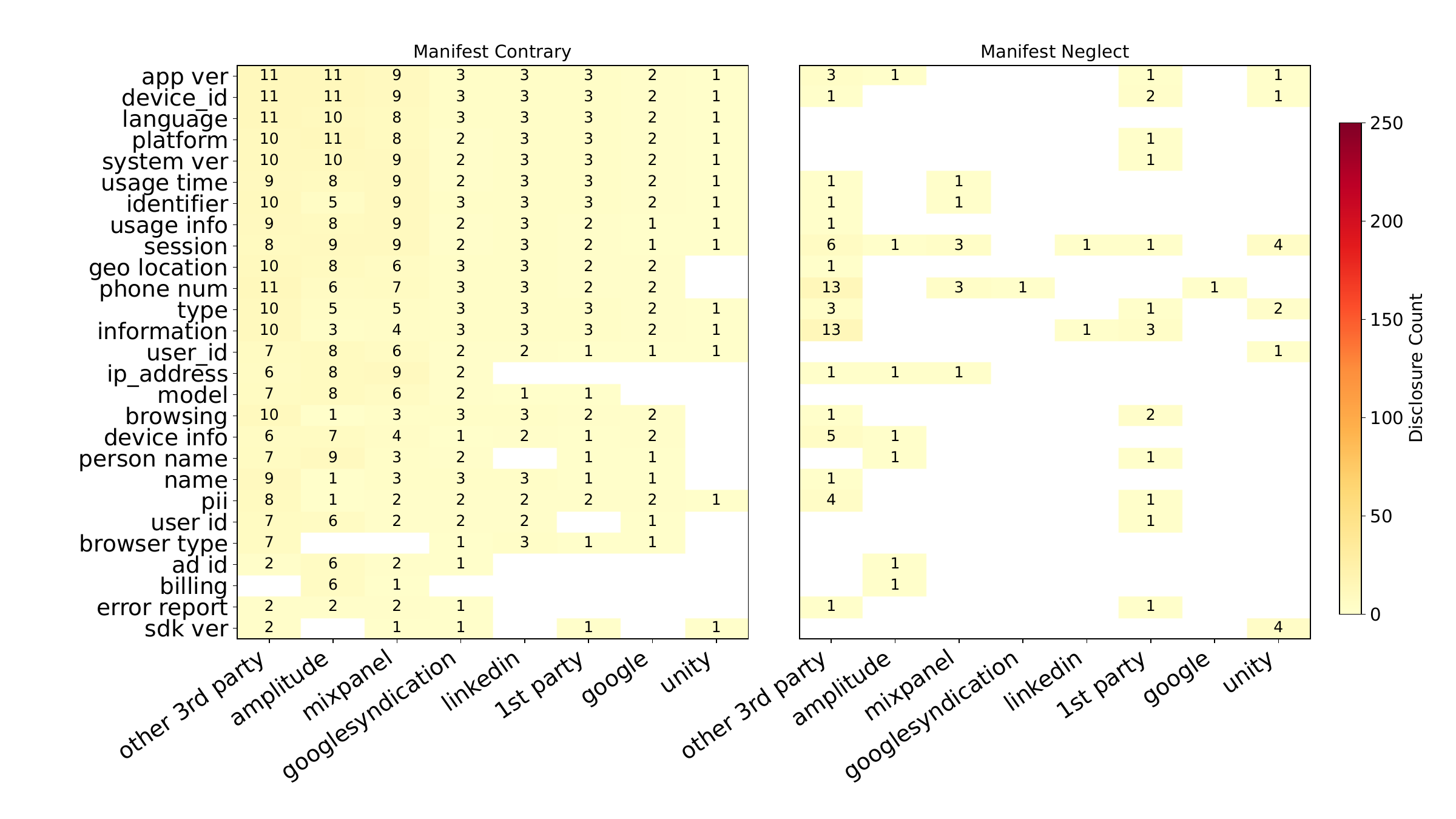}
    \caption{Results: Privacy Manifest (Partial)}
    \label{fig:violation:manifest}
\end{figure}

The heatmaps for privacy policy violation and manifest violation are presented in~\autoref{fig:violation:policy} and~\autoref{fig:violation:manifest}.

   \begin{figure}[t]
        \centering
        \includegraphics[width=.99\linewidth]{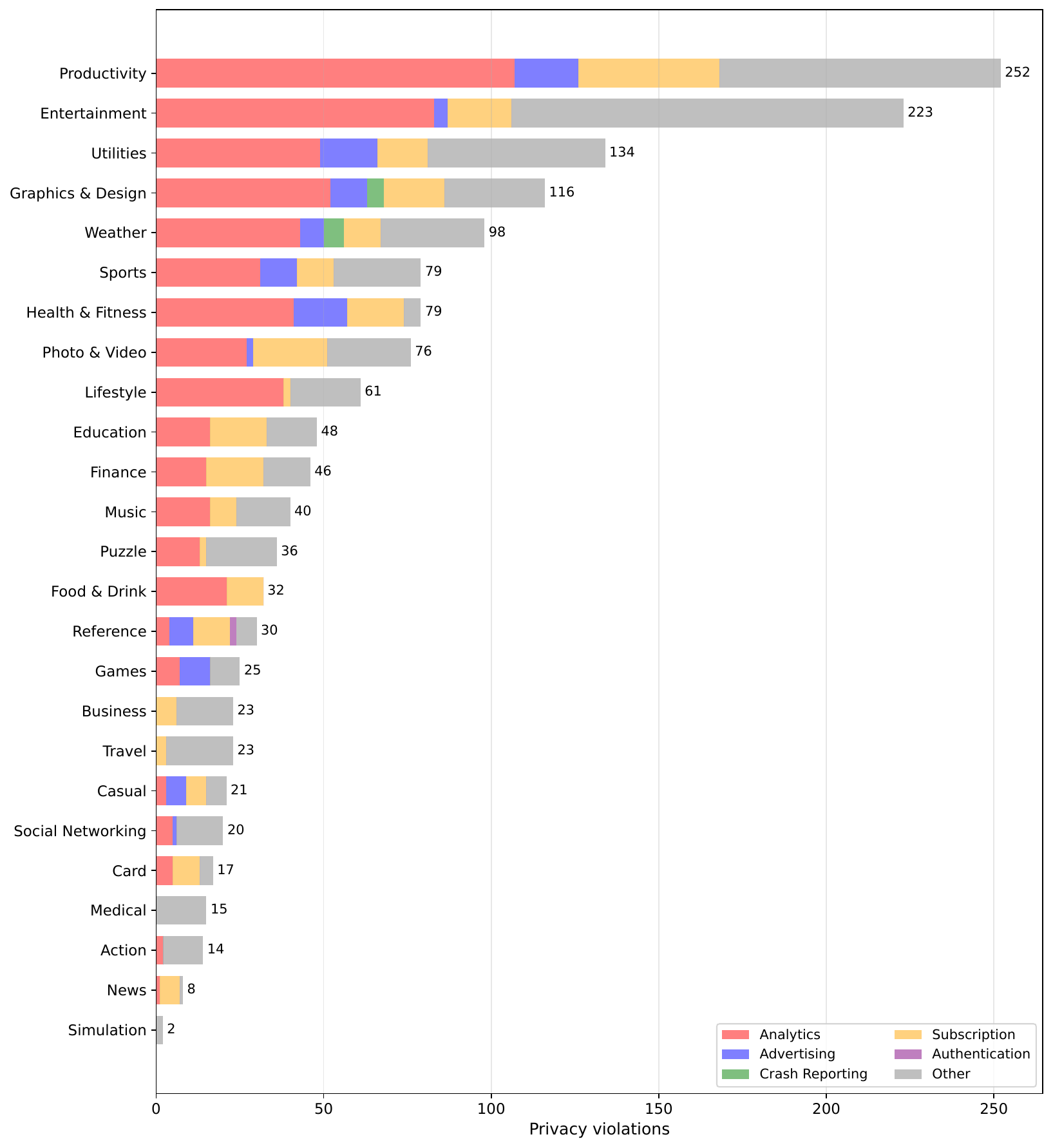}
        \caption{Privacy-policy violations  per app category w.r.t. SDKs; colors indicate the SDK category of the receiving entity.}
        \label{fig:policy-violations-by-cat-sdk}
    \end{figure}

    \begin{figure}[t]
        \centering
        \includegraphics[width=.99\linewidth]{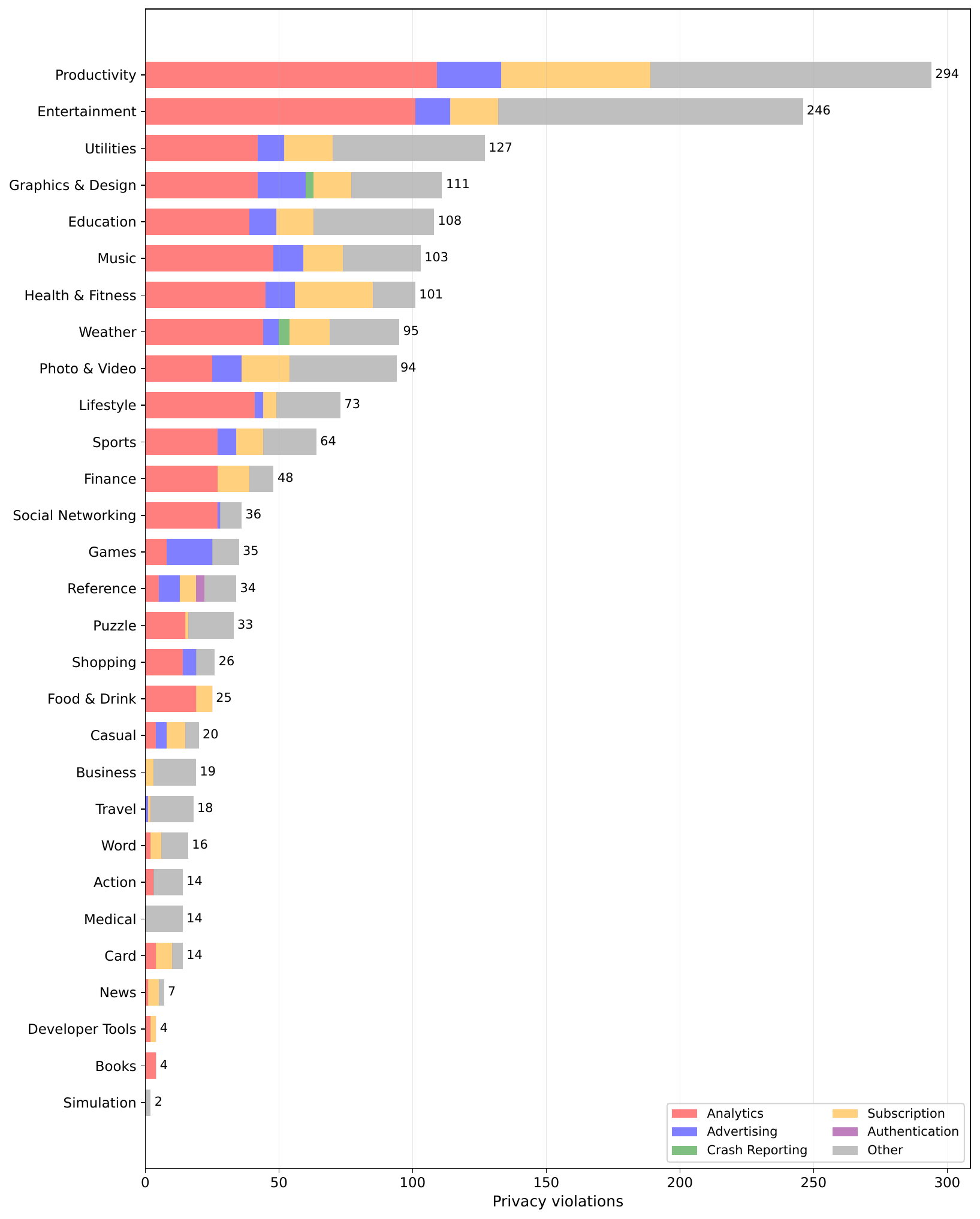}
        \caption{Privacy-label violations per app category w.r.t. SDKs; colors indicate the SDK category of the receiving entity.}
        \label{fig:label-violations-by-cat-sdk}
    \end{figure}
    
\begin{figure}[t]
    \centering
    \includegraphics[width=.99\linewidth]{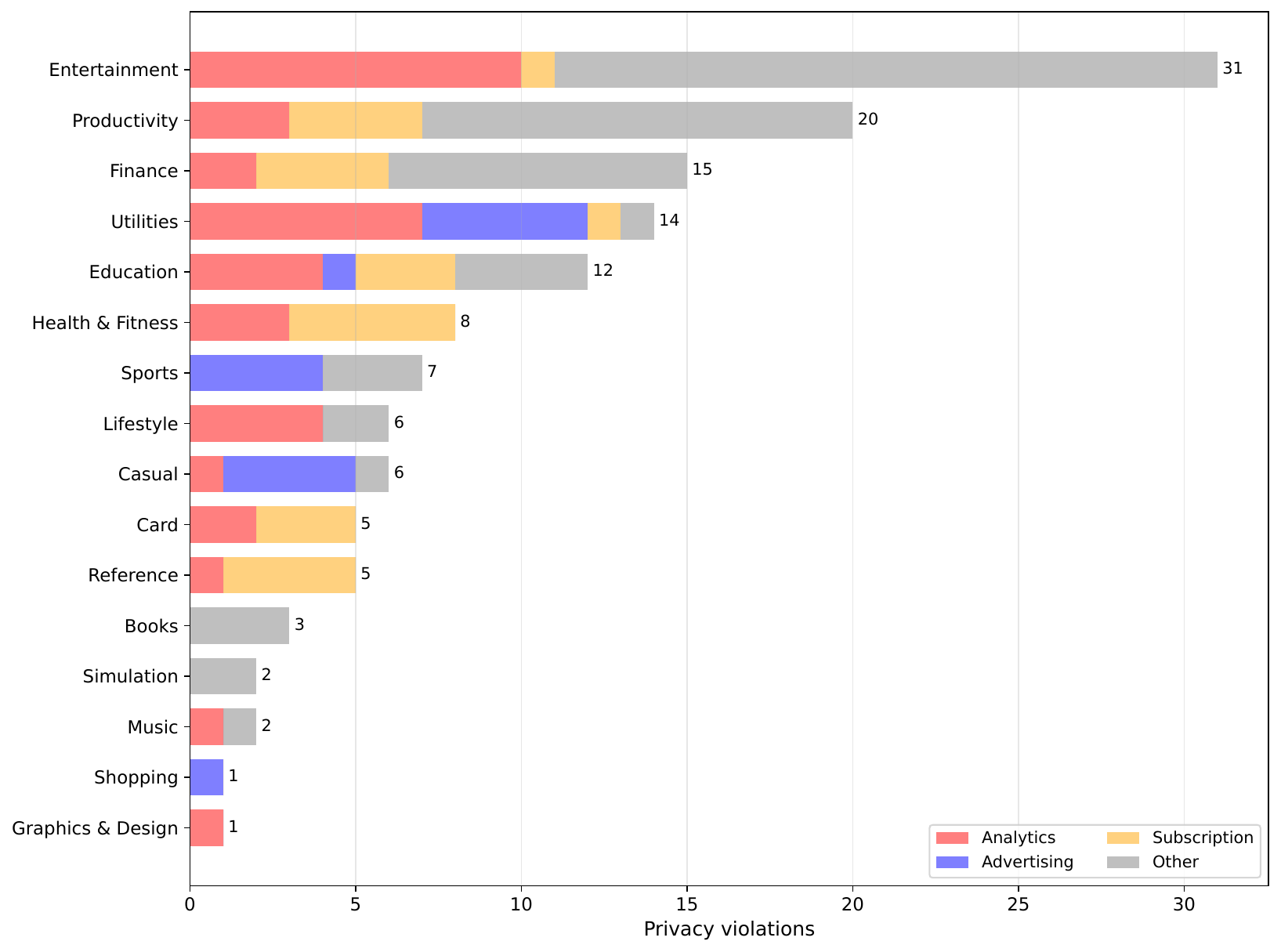}
    \caption{Privacy-manifest violations per app category w.r.t. SDKs; colors indicate the SDK category of the receiving entity.}
    \label{fig:manifest-violations-by-cat-sdk}
\end{figure}

\section{SDK Analysis}
\label{sec:app:SDK}

Figures~\ref{fig:policy-violations-by-cat-sdk},~\ref{fig:label-violations-by-cat-sdk}, ~\ref{fig:manifest-violations-by-cat-sdk}  show the SDK-related violations mentioned in~\autoref{sec:sdk}.

\section{Additional Related Work}
\label{sec:app:related}

\bheading{XR privacy attacks and defenses.}
Existing works mainly exploit machine learning models to perform side-channel attacks to breach user's privacy on XR platforms.
Researchers have  performed user identification based on motion data in XR~\cite{moore_personal_2021,nair_unique_2023,miller_personal_2020,miller_large-scale_2023,miller_using_2021,miller_within-system_2020,pfeuffer_behavioural_2019,rack_extensible_2023,liebers_exploring_2023,liebers_identifying_2024,liebers_understanding_2021,sabra_exploiting_2023,jarin_behavr_2023,meng_-anonymization_2023,garrido_sok_2024,stephenson_sok_2022,liebers_kinetic_2024}.
Besides identifying users,
motion data in XR has also been used to infer 
specific user attributes such as age and gender (\ie, user profiling)~\cite{nair_truth_2023,nair_inferring_2023}
from head and hand movement patterns.
Other secrets such as keystrokes or speech in XR
can also be inferred
from motion data~\cite{zhang_its_2023,wu_privacy_2023,luo_holologger_2022,slocum_going_2023,meteriz-yildiran_keylogging_2022,cayir2025speak,ye2024bpsniff,lee2025eyes},
videos of user movements~\cite{gopal_hidden_2023,nguyen_penetration_2024,yang_can_2023},
acoustic signals~\cite{luo_eavesdropping_2024}, network traffic~\cite{su2024remote},
infrared signals~\cite{ni2024non},
WiFi signals~\cite{al_arafat_vr-spy_2021} and GPU profiling data~\cite{son2025side}.
Aziz~\etal~\cite{aziz2025exploring} investigate how unprotected body motion data can weaken privacy safeguards for eye-tracking data, and conversely how eye-tracking data can compromise protections on motion data, ultimately facilitating user identification in VR.
Wang~\etal~\cite{wang2024gazeploit} demonstrated the first side-channel attack on AVP,
where they used eye movement information exposed from AVP Persona to infer keystrokes.
Recent works~\cite{nair_deep_2023, nair_going_2023,li_kaleidoreal-time_2021,david2023privacy} have proposed injecting noises into motion data or identifiable anthropometrics to make inference attacks harder,
which have predominantly used differential privacy.
Shoaib~\etal proposed RealityCheck~\cite{shoaib2025principled}, a provenance-based auditing system and investigated 25 XR attacks.
There are SoKs~\cite{garrido_sok_2024,stephenson_sok_2022} summarizing XR attacks and defenses.

\bheading{iOS privacy analysis.}
PiOS \cite{EgeleKKV11} applied static analysis to Mach-O binaries to track data flows, revealing that about half of 1,400 analyzed apps leaked the device's unique ID.
iRis \cite{DengSZX15} combined static and iterative dynamic analysis to detect security-critical private API usage, identifying 146 (7\%) apps that accessed sensitive data.
Other works have used source-to-sink analysis to detect cryptographic API misuse (iCryptoTracer \cite{li2015icryptotracer}) or private API usage that exposes personally identifiable information (iAnalytics \cite{zheng2015enpublic}).
Additional studies include a crowdsourcing effort \cite{agarwal2013protectmyprivacy} showing the prevalence of privacy-sensitive resource access, and a proposal for fine-grained, user-driven sandboxing for runtime privacy protection \cite{werthmann2013psios}.
Beyond invasive behavior detection, recent work has focused on privacy compliance analysis, assessing whether observed data practices align with privacy documentation such as Apple's privacy labels, privacy policies, and disclosure guidelines.
Recent studies \cite{xiao2023lalaine, kollnig2022goodbye,surma2024examining} examined the accuracy of Apple's privacy labels and their consistency with actual data handling behaviors.
Other work has focused on third-party libraries: Colaine~\cite{xiao2024measuring} leveraged NLP and dynamic analysis to verify compliance with privacy label disclosure guidelines, while iHunter~\cite{liu2024ihunter} employed static taint analysis with iOS-specific symbolic execution and NLP-based taint rule generation to identify compliance violations in the iOS software supply chain.
Unlike prior work that focuses on traditional iOS apps and SDKs, our work specifically investigates privacy risks and compliance issues in  AVP apps.

\begin{table}[t]
\centering
\footnotesize
\caption{Four apps have immersive scenes in our ground-truth dataset; our manual analysis showed that none of them generates network traffic.}
\label{tab:immersive-no-traffic}
\begin{tabular}{p{3cm}p{4.8cm}}
\toprule
\textbf{App} & \textbf{Immersive functionality} \\
\midrule
\emph{Spatial Physics Playground} & Rendering VR backgrounds \\
\emph{What If...?} & Interacting with prebuilt Disney characters \\
\emph{3D Tic Tac Toe} & Playing 3D interactive games \\
\emph{Breadpad} & Breadboard simulator \\
\bottomrule
\end{tabular}
\end{table}

\section{More Discussions}
\label{sec:app:more-discuss}

\subsection{Traffic Analysis for Immersive Scenes}
\label{sec:app:immersive}

There are four apps (\autoref{tab:immersive-no-traffic}) in our ground-truth dataset that have immersive scenes, which \sysname cannot explore due to our limitations.
However, through manual analysis on the immersive scenes in those apps,
we found that they did not generate any network traffic. 
Therefore, the immersive scenes are unlikely to impact our final results.

\subsection{Limitations}
The current design of \sysname has the following limitations.
First,
\sysname relies on the image recognition tool (\cc{OmniParserV2} in our implementation) to recognize the clickable objects, which often generates randomness and prone to errors.
Second,
while \sysname can simulate user inputs such as clicks,
it cannot simulate complex user interactions that require hand movements. As a result, we cannot handle complex immersive scenes.
However, our manual exploration showed that only 4 out of 50 ground truth apps have such scenes, and those scenes do not send any traffic (\autoref{tab:immersive-no-traffic} in   Appendix~\autoref{sec:app:immersive}).

\ycrv{Moreover, our UI-based exploration cannot create accounts or complete in-app purchases, so functionality behind such gates is unreachable regardless of exploration time. 
We manually audited the recorded sessions of all 50 ground-truth apps. 
Two apps place a login wall in front of essentially all functionality, so our traversal of those apps is limited to their pre-authentication screens. Five apps gate a subset of premium features behind an in-app purchase; because we did not transact, we cannot characterize what those paths would expose. The remaining 43 apps are fully explorable. 
Therefore, our method can already cover 48/50 apps.
}

\ycrv{Our taxonomy inherits its synonym lists verbatim from PoliCheck/VPVet,
and a few are broader than the AVP setting warrants. The clearest case is
\emph{password}, whose inherited synonym list includes the term \cc{token};
because such tokens are typically used for authentication rather than being
credentials themselves, flows matched through this term should be read as
authentication-related rather than as transmitted passwords.}

\subsection{Future Work}
\label{sec:app:future}

\bheading{Using GUI agents for exploration.}
GUI agents are very popular recently, and they have been used to automatically test Android apps~\cite{zhang2024android,ye2025mobile,yoon2024intent,zhang2025agentcpm}.
One promising direction is to replace \comptwo with a GUI agent for auto-exploration.
We have attempted to use GUI agents, but we encountered the following challenges.
First,
existing GUI agents are trained using 2D images and trajectories collected from smartphones,
which is not adaptable to AVP scenarios.
Second,
existing GUI agents often rely on debugging tools such as ADB to control the device,
which is not available on AVP.
As a result,
in this paper,
we chose not to use GUI agents.

\bheading{Other applications of \sysname.}
While this paper only presents one concrete application of using \sysname to detect privacy violations in network traffic,
\sysname (especially \compone and \comptwo) can be easily extended to other applications.
For example,
\sysname can be repositioned to perform fuzz testing on XR devices to detect functional bugs in XR applications;
it can also be used to perform performance testing for XR applications.
With the recent advancements of vision language models (VLMs),
\sysname can also be used to collect UI data and study the semantics of XR UI contents using VLMs.
\sysname can also be extend to other devices beyond AVP by replacing \compone with device-specific hardware.
We will open-source \sysname to facilitate future research.

\section{Implementation of \sysname}
\label{sec:impl}
We implement a prototype of \sysname, 
which consists of 286 lines of ESP32 code 
and about 3,000 lines of Python code.
In this section, 
we present details of our implementation.

\subsection{\compone}
\label{sec:impl:compone}

\bheading{Simulating User Inputs.}
We implement the customized HID using an Arduino ESP32-S3 microcontroller with built-in Bluetooth Low Energy support. 
We develop custom firmware in C that implements the HID profile to emulate mouse and keyboard inputs. 
The firmware adapts the connection protocol to satisfy AVP's strict device pairing requirements~\cite{Apple-accessory} and optimizes input command queuing and execution timing to match AVP's response characteristics.

\bheading{Pressing Physical Buttons.}
We select servo motors based on torque requirements, positioning accuracy, and response time. 
Through empirical measurement, we determine that approximately 350g of force is required to press each physical button on the AVP device. 
We therefore select servo motors capable of providing equivalent force.
To ensure system stability during automated operation, we design a 3D-printed bracket that mounts the servo motors onto the side arms of the AVP headset. 
The bracket securely holds the servo motors without interfering with normal headset operation.
We integrate the servo motors with the ESP32 through GPIO pins with proper power management circuitry, ensuring reliable button actuation.

\subsection{\comptwo}
\label{sec:impl:comptwo}

\bheading{Screen Capturer.}
For screen capture, we configured AirPlay mirroring with optimal resolution and frame rate settings to project the AVP screen to a Macbook Pro. We use the MacBook Pro's built-in screen recording APIs with programmatic control. 
For evaluation purposes, we actually recorded the entire exploration process,
and we take screenshots  for every action by \comptwo to process.
Our frame processing pipeline goes from capture to format conversion to analysis preparation. We optimized latency to balance capture quality with real-time interaction requirements.

\bheading{Virtual Cursor Handler.}
We locate the virtual cursor in screenshots using a two-step computer vision pipeline.
First, we apply an HSV color filter with hue range $[36, 86]$ to isolate the green cursor from the background, effectively removing noise from the scene.
Second, we detect circular shapes using the \textit{Hough Circle Transform}~\cite{OpenCV_Circle_Hough_Transform} with parameters \texttt{minRadius=20} and \texttt{maxRadius=100}.
To handle variations in cursor size, we employ a progressive matching strategy that gradually relaxes the radius tolerance from 20\% to 80\%.
A key challenge arises from AVP's pointer control system, which restricts cursor movement to the boundaries of the currently focused window panel.
Since cursor movement is only permitted within the bounds of the focused window, we must ensure proper window focus before attempting UI interactions.
We address this by leveraging AVP's head tracking feature to center the headset on the target window, which causes newly opened windows to appear at the center of the 3D scene and automatically receive focus.
This approach enables unrestricted cursor movement within the focused window boundaries.

\bheading{State Explorer.}
To construct UI states from screenshots, we deploy a pretrained OmniParserV2 model~\cite{omniparserv2}. 
The model inference runs on a GPU server through a REST API wrapper, which the MacBook queries directly for each screenshot.
We configure a detection confidence threshold of 0.05 to ensure accurate button detection in 3D environments. 
For state comparison, we extract text labels from interactive UI elements detected by OmniParser and compute pairwise text similarity using normalized Levenshtein distance with OCR-confusion correction.
Two states are considered equivalent when (1)~their button counts differ by less than 30\%, and (2)~the average best-match text similarity exceeds 0.7.
A key challenge arises when the exploration becomes trapped in complex 3D scenes that require hand gestures and head movements for interaction.
Our manual analysis reveals that these immersive scenes typically do not trigger network traffic.
To prevent indefinite exploration of such states, we implement a timeout detector that monitors UI state changes.
When the UI remains unchanged for more than one minute, we mark the current state as \textit{invalid} and invoke the ``Force Quit'' feature from \compone to terminate the application.
Subsequently, any exploration paths leading to \textit{invalid} states are excluded from future iterations.

\subsection{\compthree}
\label{sec:impl:compthree}
\bheading{Network Traffic Capture.}
We capture network traffic from the AVP device using a Man-in-the-Middle (MITM) proxy approach.
Specifically, we deploy \cc{Mitmproxy} on the MacBook and configure the AVP device to connect through the MacBook's Wi-Fi hotspot.
To enable HTTPS traffic decryption, we install \cc{Mitmproxy}'s root certificate on the AVP device  
\ycrv{and mark it as fully trusted in visionOS's Certificate Trust Settings}.
This setup allows us to intercept, decrypt, and analyze all network communications originating from the AVP device during application testing.

\section{Detector Details}
\label{app:detector}

\ycrv{\textbf{Matching mechanisms.} 
We map extracted keys to taxonomy nodes using three mechanisms. \emph{Synonym matching} uses the synonym lists of our taxonomy; 
these are inherited from VPVet and PoliCheck~\cite{zhan2024vpvet,andow2020actions}, then
augmented with the phrase groups extracted from network traffic. \emph{Key-pattern matching} uses a manually curated dictionary of common SDK/API field names (e.g., \cc{lat}/\cc{lon}) and hand-written key regular expressions, extracted from the field names we observed across the collected traffic. 
\emph{Value-pattern matching} uses manually defined regular expressions that recognize canonical value formats: email addresses, IP addresses, UUIDs, and latitude/longitude pairs.}

\ycrv{\textbf{First-party vs.\ third-party attribution.}
Following OVRSeen~\cite{trimananda2022ovrseen}, we classify a flow as
first-party when its destination domain matches the app's own domain, a domain
declared in its privacy manifest, or the domain hosting its privacy policy. All
remaining destinations default to third-party; we manually labelled the most
frequently observed third-party entities and maintain that list as part of our
artifact. Because a first-party backend hosted on an independently registered
domain still defaults to third-party, our first-party counts are a lower bound.}

\ycrv{\textbf{Why the disclosures are not ground truth.}
Neither the privacy policy nor the privacy label can serve as ground truth:
both are developer-provided disclosures and can be incomplete or incorrect, and
ground truth can only be obtained from app behavior. Our work therefore
compares observed runtime data collection against each privacy document
separately. Pure policy--label inconsistency is orthogonal to our goal and can
be studied with existing privacy-document analysis tools such as PolicyLint~\cite{andow2019policylint}.}

\section{Long-running Exploration}
\label{app:longrun}

\ycrv{To assess both how much \sysname captures and how stably it behaves over
time, we re-ran 3 apps from different category with a 60-minute budget.}

\begin{figure}[t]
  \centering
  \includegraphics[width=\linewidth]{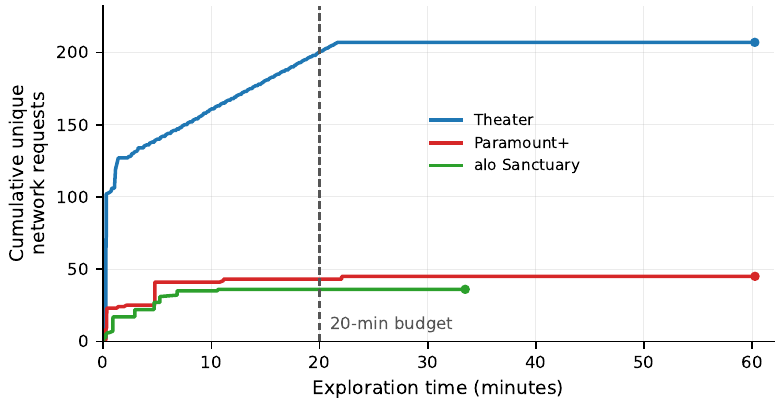}
  \caption{\ycrv{Cumulative unique network requests during extended 60-minute sessions. All three apps saturate before the 20-minute budget expires.}}
  \label{fig:saturation}
\end{figure}

\ycrv{\bheading{Coverage.} In all three, the cumulative number of unique requests
saturates well before the 20-minute budget expires (\autoref{fig:saturation}):
96.6\%, 95.6\%, and 100\% of the requests seen over the full session are
already present at 20 minutes.}

\ycrv{\bheading{Reliability.} The plateau reflects saturation of the reachable
surface rather than the explorer stalling. \cc{Theater} keeps interacting
throughout the remaining 39 minutes---243 clicks across 124 states over the
full hour---while discovering no new request after minute~21, and a second
60-minute session issued 217 clicks across 33 states without interruption.
Exploration therefore does not degrade over time.
}

\section{TLS Interception Coverage}
\label{app:tls}

\ycrv{We re-examined the raw \cc{mitmproxy} captures for the full large-scale
dataset to determine whether any traffic escaped interception.}

\ycrv{\bheading{Decryption.} Every intercepted connection completed the TLS handshake against our proxy certificate. We observed no certificate rejections and no undecryptable payloads, and all 336 distinct destination hosts yielded successfully decrypted requests.}

\ycrv{\bheading{Errored flows.} Of the 201 errored flows, 165 originate from transient failures on the local proxy hop and the remainder are ordinary mid-stream resets. None exhibits the behavior characteristic of certificate pinning, i.e., a handshake refused with a certificate error, or a connection established but immediately dropped before any request is sent.
}

\ycrv{\bheading{What cannot be observed.} The one behavior invisible on-path is silent evasion, in which an app declines to connect at all, or routes around the proxy, and therefore leaves no flow to analyze; detecting it would require observing runtime symptoms such as error dialogs or broken functionality. 
Our screen recordings show no such symptoms, but because silent evasion is undetectable by construction, we treat any flow we cannot decrypt as unknown rather than benign and present our measurements as a lower bound.}

\subsection{Case Studies}
\label{sec:case-studies}
\label{app:case-studies}

We present two representative case studies that illustrate the privacy disclosure violations discovered in our analysis. The first case demonstrates a severe \textit{Contrary Disclosure} violation where an app explicitly claims to collect no data while actually collecting extensive user information. The second case illustrates a \textit{Neglect Disclosure} violation in a Vision Pro-specific spatial computing application.

\vspace{3pt}
\noindent\textbf{\textit{Contrary Disclosure (AirLauncher App Launcher).}}
AirLauncher~\cite{Airlaucher} is a utilities app designed exclusively for AVP. The app serves as a shortcut widget that allows users to quickly launch websites, apps, contacts, and shortcuts with a single pinch gesture.
Critically, the app's App Store privacy label explicitly declares \texttt{``Data Not Collected''}, stating: \textit{``The developer does not collect any data from this app.''} However, our dynamic analysis reveals that the app transmits \textit{9 distinct privacy-related data items} to RevenueCat's subscription management API, 
including \textit{Device Identifier}, \textit{App Fingerprint}, \textit{Usage Behavior}, and \textit{Performance Metrics}. 
RevenueCat, a third-party analytics SDK,  can support cross-app attribution and identifier linkage, which increases the risk of user profiling and tracking and directly contradicts the app’s stated disclosure.

\ignore{
The collected data includes:
\begin{itemize}
    \item \textbf{Device Identifier}: The unique Apple device identifier (\texttt{x-apple-device-identifier}: \texttt{E00CCD7B-162E-...}) enabling persistent cross-session user tracking.
    \item \textbf{Platform Information}: Operating system (\texttt{visionOS}), version (\texttt{Version 2.1 Build 22N581}), and platform flavor (\texttt{native}).
    \item \textbf{App Fingerprint}: Bundle ID (\texttt{co.swiftfox.LaunchBar}), client version (\texttt{1.5.5}), build version (\texttt{100}), and SDK version (\texttt{4.39.0}).
    \item \textbf{Usage Behavior}: StoreKit interaction flags (\texttt{x-storekit2-enabled}) and observer mode settings.
    \item \textbf{Performance Metrics}: Service response times via \texttt{x-envoy-upstream-service-time} headers (values: 3ms, 5ms, 451ms).
\end{itemize}
}

\vspace{3pt}
\noindent\textbf{\textit{Neglect Disclosure (Twin: Scans for Passthrough).}}
Twin~\cite{Twin} is a spatial computing productivity app that exemplifies privacy concerns unique to the AVP ecosystem. 
The app utilize visionOS capabilities including photogrammetry scanning, passthrough visualization, and immersive environment placement to create \emph{digital twins} of physical objects.
Users scan real-world items (such as keyboards or desk decorations) using their iPhone or iPad, then place these 3D models in Vision Pro's virtual environments, enabling them to see physical objects while fully immersed.
This use case raises heightened privacy concerns specific to Vision Pro: users scan their physical surroundings and synchronize data across multiple devices (iPhone/iPad for scanning, Vision Pro for viewing), creating a multi-device data trail that could reveal information about users' physical spaces and daily activities.
The app's privacy label declares only \emph{Data Not Linked to You} without specifying categories, leaving users unaware of the extent of data collection. Our analysis reveals \textit{8 undisclosed data collection practices} spanning 4 privacy categories, including \textit{Identifiers}, \textit{Location}, \textit{Usage Data} and \textit{Diagnostics}, sending to two third-party SDKs: RevenueCat (subscription management) and Paywalls (UI presentation). 
The visionOS-specific headers in the traffic (\texttt{x-platform: visionOS} and \texttt{x-platform-version: Version 2.1}) explicitly identify the device as Apple Vision Pro, enabling platform-specific user profiling. 
This case highlights a growing Vision Pro privacy issue: as apps use spatial computing features, sensitive data collection rises, but privacy disclosures often overlook platform-specific risks and third-party SDK behavior.
\looseness=-1

\ignore{
\begin{itemize}
    \item \textbf{Identifiers}: Device identifier (\texttt{x-apple-device-identifier}: \texttt{DDCFF958-F838-...}) and user authorization token sent to \texttt{api.revenuecat.com}.
    \item \textbf{Location}: Geographical location inferred from CDN edge node (\texttt{x-amz-cf-pop}: \texttt{IAD89-P1}, indicating Washington D.C. area) via \texttt{assets.pawwalls.com}.
    \item \textbf{Usage Data}: Store interaction flags (\texttt{x-storekit2-enabled}) and usage patterns.
    \item \textbf{Diagnostics}: Performance timing data (\texttt{x-envoy-upstream-service-time}: 3ms, 5ms, 67ms) sent to RevenueCat servers.
\end{itemize}
}

\end{document}